\documentclass[final,3p,times,twocolumn]{elsarticle}
\usepackage[frozencache,cachedir=.,newfloat]{minted} 

\usepackage{subcaption}
\usepackage[newfloat]{minted} 
\usepackage{hyperref}
\usepackage{tcolorbox}
\usepackage{etoolbox}
\BeforeBeginEnvironment{minted}{\begin{center}\begin{tcolorbox}[width=\linewidth]}
\AfterEndEnvironment{minted}{\end{tcolorbox}\end{center}}
\usepackage{caption}
\usepackage{xspace}
\usepackage[toc,page]{appendix}

\newcommand{\pvec}[1]{\vec{#1}\mkern2mu\vphantom{#1}}
\newcommand{\inline}[1]{\mintinline{python}{#1}\xspace}
\newcommand{\SuperScreen}{\inline{SuperScreen}}
\newcommand{\um}{\mu\mathrm{m}}
\newcommand{\nm}{\mathrm{nm}}

\newenvironment{code}{\captionsetup{type=listing}}{\hfill}
\SetupFloatingEnvironment{listing}{name=Code Block}
\BeforeBeginEnvironment{code}{\vspace{10pt}\begin{figure*}[t!]}
\AfterEndEnvironment{code}{\end{figure*}}
\newenvironment{code-onecol}{\captionsetup{type=listing}}{\hfill}
\SetupFloatingEnvironment{listing}{name=Code Block}
\BeforeBeginEnvironment{code-onecol}{\vspace{10pt}\begin{figure}[t!]}
\AfterEndEnvironment{code-onecol}{\end{figure}}

\usepackage{amsmath, amssymb, bm}

\newcounter{bla}

\journal{Computer Physics Communications}

\begin{document}

\begin{frontmatter}



\title{\SuperScreen: An open-source package for simulating the magnetic response of two-dimensional superconducting devices}


\author[PHYS,SIMES]{Logan Bishop-Van Horn\corref{lbvh}}\ead{lbvh@stanford.edu}
\author[PHYS,AP,SIMES]{Kathryn A. Moler}\ead{kmoler@stanford.edu}

\cortext[lbvh]{Corresponding author.}

\address[PHYS]{Department of Physics, Stanford University, Stanford, California 94305, USA}
\address[AP]{Department of Applied Physics, Stanford University, Stanford, California 94305, USA}
\address[SIMES]{Stanford Institute for Materials and Energy Sciences, SLAC National Accelerator Laboratory, 2575 Sand Hill Road, Menlo Park, California 94025, USA}

\begin{abstract}
Quantitative understanding of the spatial distribution of magnetic fields and Meissner screening currents in two-dimensional (2D) superconductors and mesoscopic thin film superconducting devices is critical to interpreting the results of magnetic measurements of such systems. Here, we introduce \SuperScreen, an open-source Python package for simulating the response of 2D superconductors to trapped flux and applied time-independent or quasi-DC magnetic fields for any value of the effective magnetic penetration depth, $\Lambda$. Given an applied magnetic field, \SuperScreen solves the 2D London equation using an efficient matrix inversion method~\cite{Brandt2004-ew, Brandt2005-wj} to obtain the Meissner currents and magnetic fields in and around structures composed of one or more superconducting thin films of arbitrary geometry. \SuperScreen can be used to model screening effects and calculate self- and mutual-inductance in thin film superconducting devices.
\end{abstract}


\end{frontmatter}


\noindent
{\bf PROGRAM SUMMARY}

\begin{small}
\noindent
{\em SuperScreen}\\
{\em CPC Library link to program files:} (to be added by Technical Editor) \\
{\em Developer's repository link:} \href{http://www.github.com/loganbvh/superscreen}{www.github.com/loganbvh/superscreen}\\
{\em Code Ocean capsule:} (to be added by Technical Editor)\\
{\em Licensing provisions:} \href{https://opensource.org/licenses/MIT}{MIT License}\\
{\em Programming language:} Python\\
{\em Nature of problem:} \SuperScreen solves for Meissner screening currents in structures composed of 2D or thin film superconductors in the presence of an applied magnetic field, pinned vortices, and trapped flux.\\
{\em Solution method:} This package solves the 2D London equation for superconducting thin films using a matrix inversion method~\cite{Brandt2004-ew,Brandt2005-wj}.\\
   \\

\end{small}

\section{Introduction}
\label{section:introduction}

\SuperScreen is a Python package developed to simulate the static magnetic response of structures composed of one or more layers containing superconducting thin films characterized by a London penetration depth $\lambda$ that is large compared to the film thickness $d$. \SuperScreen solves the coupled Maxwell's and London's equations in and around superconducting films in the presence of inhomogeneous applied magnetic fields, pinned vortices, and trapped flux using a matrix inversion method introduced by Brandt and Clem~\cite{Brandt2004-ew,Brandt2005-wj} and subsequently used by Kirtley, \textit{et al.} to model the magnetic response of scanning superconducting quantum interference device (SQUID) sensors~\cite{Kirtley2016-zz, Kirtley2016-gt}.

There have been many previous numerical studies of magnetic screening and inductance extraction in thin film and two-dimensional (2D) superconducting devices~\cite{Jaycox1981-zl, Ketchen1982-at, Ketchen2012-mb, Hildebrandt1995-uw, Khapaev1997-kw, Khapaev2001-xq, Khapaev2001-pw, Khapaev2002-ev,  Babaei_Brojeny2003-la, Brandt2004-ew, Brandt2005-wj, Clem2005-ye, Muller2021-ci, Fourie2011-wl, Fourie2012-gv, Jackman2016-mf}. However, few software tools for this task exist and those that are available are closed-source, are written in low-level compiled languages like C, and/or require the use of specialized file formats or separate computer aided design (CAD) software for defining device geometries and model configurations.\footnote{A brief summary of existing tools can be found in \ref{appendix:other-tools}.} Most of these tools are intended for use in the design of superconducting integrated circuits for single flux quantum (SFQ) logic and are primarily used for inductance extraction~\cite{Gaj1999-ls, Tolpygo2021-jz}. \SuperScreen is an open-source, user-friendly, portable research tool designed to lower the barrier to entry to quantitative modeling of 2D superconductors, and to help interpret and inform measurements of superconducting thin films and devices.

This introduction to \SuperScreen is organized as follows: In Section~\ref{section:model} we outline the model and its assumptions, and in Section~\ref{section:implementation} we describe its numerical implementation. In Section~\ref{section:overview} we provide an overview of the structure of the \SuperScreen package and discuss some important development details. In Section~\ref{section:examples} we demonstrate how to perform several types of simulations using \SuperScreen and compare the results to analytical solutions and experimental results. Finally, in Section~\ref{section:conlusion} we conclude by discussing applications, limitations, and possible extensions of the package.

\section{The Model}
\label{section:model}

The goal of \SuperScreen is to model the magnetic response of a thin superconducting film, or a structure composed of multiple superconducting films (which may or may not lie in the same plane), to an applied inhomogeneous out-of-plane magnetic field
$H_{z,\,\mathrm{applied}}(x, y, z)$. Given $H_{z,\,\mathrm{applied}}(x, y, z)$ and information about the geometry and magnetic penetration depth of all films in a superconducting structure, we aim to calculate the thickness-integrated current density $\vec{J}(x, y)$ at all points inside the films, from which one can calculate the vector magnetic field $\vec{H}(x, y, z)$ at all points both inside and outside the films.

A convenient method for solving this problem was introduced by Brandt and Clem in Ref.~\cite{Brandt2004-ew}, expanded by Brandt in Ref.~\cite{Brandt2005-wj}, and subsequently used to model the magnetic response of scanning SQUID susceptometers~\cite{Kirtley2016-zz, Kirtley2016-gt}. In the London model of superconductivity, the magnetic field $\vec{H}(\vec{r})$ and 3D current density $\vec{j}(\vec{r})$ in a superconductor with London penetration depth $\lambda$ obey the second London equation:
$\vec{H}(\vec{r})/\lambda^2=-\vec{\nabla}\times\vec{j}(\vec{r})$, where
$\vec{\nabla}=\left(\frac{\partial}{\partial x}, \frac{\partial}{\partial y}, \frac{\partial}{\partial z}\right)$ is the 3D gradient operator. The 2D London model assumes that the current density $\vec{j}$ is approximately independent of $z$, such that  $\vec{j}(\vec{r}) = \vec{j}(x, y, z)\approx\vec{j}_{z_0}(x, y)$ for a film lying parallel to the $x-y$ plane at vertical position $z_0$. Working now with the thickness-integrated current density $\vec{J}(x, y)=\vec{j}_{z_0}(x, y)\cdot d$, where $d$
is the thickness of the film, the second London equation reduces to

\begin{equation}
    \label{eq:london}
    \vec{H}(x, y)=-\Lambda\vec{\nabla}\times\vec{J}(x, y)
\end{equation}
where $\Lambda=\lambda^2/d$ is the effective penetration depth
of the superconducting film (equal to half the Pearl length~\cite{Pearl1964-cl}) and $\vec{\nabla}=\left(\frac{\partial}{\partial x}, \frac{\partial}{\partial y}\right)$ is now the 2D gradient operator.

It is important to note that the assumption $\vec{j}(x, y, z)\approx\vec{j}_{z_0}(x, y)$ is valid for only films that are thinner than their London penetration depth ($d\ll\lambda$, such that $\Lambda=\lambda^2/d\gg\lambda$). However the model has been applied with some success in structures with $\lambda\lesssim d$, for example by Kirtley, \textit{et al.} in modeling the magnetic response of scanning SQUID susceptometers~\cite{Kirtley2016-zz,Kirtley2016-gt}. Aside from this limitation, the method described below can be used to model films with any effective penetration depth $0\leq\Lambda<\infty$.

Because the current density has zero divergence inside the superconducting film ($\nabla\cdot\vec{J}=0$)
except at small terminals where current can be injected, one can express $\vec{J}$ in terms
of a scalar potential $g(x, y)$, called the stream function:
\begin{equation}
    \label{eq:stream}
    \vec{J}(x, y) = -\hat{z}\times\vec{\nabla}g
    = \vec{\nabla}\times(g\hat{z})
    = \left(\frac{\partial g}{\partial y}, -\frac{\partial g}{\partial x}\right).
\end{equation}

The stream function $g$ can be thought of as the local magnetization of the film, or the area density of magnetic dipole sources (see Ref.~\cite{Brandt2005-wj} for more interesting properties of the stream function). We can rewrite Eq.~\ref{eq:london}, which gives the magnetic field inside of a 2D film, in terms of $g$:
\begin{align}
    \label{eq:london_stream}
    \begin{split}
        \vec{H}(x, y) &= -\Lambda\left[\nabla\times\vec{J}(x, y)\right]\\
        &= -\Lambda\left[\vec{\nabla}\times\left(\vec{\nabla}\times(g\hat{z})\right)\right]\\
        &= -\Lambda\left[\vec{\nabla}(\vec{\nabla}\cdot(g\hat{z}))-\nabla^2(g\hat{z})\right]\\
        &=\Lambda\nabla^2g(x,y)\hat{z},
    \end{split}
\end{align}
where $\nabla^2=\vec{\nabla}\cdot\vec{\nabla}$ is the Laplace operator. (The last line follows from the fact that $\vec{\nabla}\cdot\left[g(x,y)\hat{z}\right] = 0$.) From Ampere's Law, the three components of the magnetic field $\vec{H}(\vec{r})$ at position $\vec{r}=(x, y, z)$ due to a sheet of current lying in the $x-y$ plane (at vertical position $z'$) with stream function $g(x', y')$ are given by:
\begin{align}
    \label{eq:field_from_kernel}
    \begin{split}
        H_x(\vec{r}) &= \int_F Q_x(\vec{r},\pvec{r}')g(x', y')\,\mathrm{d}^2r'\\
        H_y(\vec{r}) &= \int_F Q_y(\vec{r},\pvec{r}')g(x', y')\,\mathrm{d}^2r'\\
        H_z(\vec{r}) &= H_{z,\,\mathrm{applied}}(\vec{r})
        + \int_F Q_z(\vec{r},\pvec{r}')g(x', y')\,\mathrm{d}^2r'.  
    \end{split}
\end{align}

Here we assume a static out-of-plane applied magnetic field $\vec{H}_\mathrm{applied}(\pvec{r}')=H_\mathrm{z,\,\mathrm{applied}}(\pvec{r}')\hat{z}$. $F$ is the film area (with $g = 0$ outside of the film), and $Q_x(\vec{r},\pvec{r}')$, $Q_y(\vec{r},\pvec{r}')$, and $Q_z(\vec{r},\pvec{r}')$ are dipole kernel functions which give the respective component of the magnetic field at position $\vec{r}=(x, y, z)$ due to a dipole of unit strength at position $\pvec{r}'=(x', y', z')$:
\begin{align}
    \label{eq:kernels}
    \begin{split}
        Q_x(\vec{r}, \pvec{r}') &=  3\frac{(x-x')(z-z')}
        {4\pi[(z-z')^2+\rho^2]^{5/2}}\\
        Q_y(\vec{r}, \pvec{r}') &=  3\frac{(y-y')(z-z')}
        {4\pi[(z-z')^2+\rho^2]^{5/2}}\\
        Q_z(\vec{r}, \pvec{r}') &=  \frac{2(z-z')^2-\rho^2}
        {4\pi[(z-z')^2+\rho^2]^{5/2}},
    \end{split}
\end{align}
where $\rho=\sqrt{(x-x')^2 + (y-y')^2}$. Eq.~\ref{eq:field_from_kernel} can also be seen as the Biot-Savart Law formulated in terms of the stream function $g$.

Comparing Eq.~\ref{eq:london_stream} and Eq.~\ref{eq:field_from_kernel}, we have in the plane of the film:
\begin{align}
    \label{eq:integral_equation}
    \begin{split}
        &\underbrace{\vec{H}(\vec{r})\cdot\hat{z} = H_z(\vec{r})
        = \Lambda\nabla^2g(\vec{r})}_{z-\text{component of the total field}}
        = \\ & \underbrace{H_{z,\,\mathrm{applied}}(\vec{r})}_{\text{applied field}}
        + \underbrace{\int_F Q_z(\vec{r},\pvec{r}')g(\pvec{r}')\,\mathrm{d}^2r'}_{\text{screening field}},
    \end{split}
\end{align}
where now $\vec{r}$ and $\pvec{r}'$ are 2D vectors, i.e. $z-z'=0$ since the film is in the same plane as itself. From Eq.~\ref{eq:integral_equation}, we arrive at an integral equation relating the stream function $g$ for points inside the superconductor to the applied field $H_{z,\,\mathrm{applied}}$:
\begin{align}
\begin{split}
    \label{eq:applied_to_stream}
    &H_{z,\,\mathrm{applied}}(\vec{r})
    = \\ &-\int_F\left[
        Q_z(\vec{r},\pvec{r}')-\delta(\vec{r}-\pvec{r}')\Lambda\nabla^2\right
    ]g(\pvec{r}')\,\mathrm{d}^2r',
\end{split}
\end{align}
where $\delta$ is the 2D Dirac delta function.

The goal, then, is to solve (invert) Eq.~\ref{eq:applied_to_stream} for a given $H_{z,\,\mathrm{applied}}$ and film geometry $F$ to obtain $g$ for all points inside the film (with the boundary condition $g=0$ enforced outside the film). Once $g(\vec{r})$ is known, the full vector magnetic field $\vec{H}(\vec{r})$ can be calculated at any point $\vec{r}$
from Eqs.~\ref{eq:field_from_kernel} and \ref{eq:kernels}.

\subsection{Films with holes}
\label{section:model:holes}

In films that have holes (regions of vacuum completely surrounded by superconductor), each hole $h$ can contain a trapped flux associated a current $I_{\mathrm{circ},\,h}$ circulating around the hole. The applied field that would cause such a circulating current is given by Eq.~\ref{eq:applied_to_stream} if we set $g=I_{\mathrm{circ},\,h}$ for all points lying inside hole $h$:
\begin{align}
\begin{split}
    \label{eq:Heff}
    &H_{z,\,\mathrm{eff},\,h}(\vec{r}) = \\
    &-\int_{\mathrm{hole}\,h}[
        Q_z(\vec{r},\pvec{r}')-\delta(\vec{r}-\pvec{r}')\Lambda\nabla^2
    ] I_{\mathrm{circ},\,h} \,\mathrm{d}^2r'.   
\end{split}
\end{align}

In this case, we modify the left-hand side of Eq.~\ref{eq:applied_to_stream} as follows:
\begin{align}
\begin{split}
    \label{eq:Heff_sub}
    &H_{z,\,\mathrm{applied}}(\vec{r}) - \sum_{\mathrm{holes}\,h} H_{z,\,\mathrm{eff},\,h}(\vec{r})
    = \\&-\int_F\left[
        Q_z(\vec{r},\pvec{r}')-\delta(\vec{r}-\pvec{r}')\Lambda\nabla^2\right
    ]g(\pvec{r}')\,\mathrm{d}^2r'.
\end{split}
\end{align}

The circulating current $I_\mathrm{circ,\,h}$ is defined as the total current crossing any curve that connects the interior of the hole $h$ (where $g=I_\mathrm{circ,\,h}$ to the exterior of the film (where $g=0$)~\cite{Khapaev2001-pw,Brandt2005-wj}.

\subsection{The fluxoid}
\label{section:model:fluxoid}

The fluxoid $\Phi^f_S$ for a 2D region $S$ with 1D boundary $\partial S$ is given by the sum of magnetic flux through $S$ and the line integral of the supercurrent density $\vec{J}$ around $\partial S$~\cite{Brandt2005-wj,Clem2005-ye,Tinkham2004-zn}:
\begin{equation}
    \Phi^f_S = \underbrace{\int_S\mu_0H_z(\vec{r})\,\mathrm{d}^2r}_\text{``flux part''} + \underbrace{\oint_{\partial S}\mu_0\Lambda\vec{J}(\vec{r})\cdot\mathrm{d}\vec{r}}_\text{``supercurrent part''}.
    \label{eq:fluxoid}
\end{equation}

The fluxoid vanishes for a region $S$ completely contained within a superconducting film that contains no holes or vortices, and has the same value for any region containing a given hole or collection of vortices in a superconducting film. This path-independence of the fluxoid follows from the static London equation (Eq.~\ref{eq:london}) on which the present model is based. Fluxoid quantization---the requirement that the fluxoid $\Phi^f_S=n\Phi_0$ where $n$ is an integer and $\Phi_0=h/2e$ is the magnetic flux quantum---is not automatically enforced by Eq.~\ref{eq:london} for multiply-connected films, however it can be included as an external constraint.

\subsection{Vortices}
\label{section:model:vortices}
In addition to being trapped in holes (see Section~\ref{section:model:holes}), flux may be trapped in a superconducting film in the form of vortices. The presence of vortices trapped in a film at positions $\vec{r}_v$ modifies Eq.~\ref{eq:Heff_sub} as follows:

\begin{align}
\begin{split}
    \label{eq:applied_to_stream_vortices}
    & H_{z,\,\mathrm{applied}}(\vec{r}) - \sum_{\mathrm{holes}\,h} H_{z,\,\mathrm{eff},\,h}(\vec{r}) - \sum_{\mathrm{vortices}\,v}\frac{\Phi_v}{\mu_0}\delta(\vec{r}-\vec{r}_v)
    = \\
    &-\int_F\left[
        Q_z(\vec{r},\pvec{r}')-\delta(\vec{r}-\pvec{r}')\Lambda\nabla^2\right
    ]g(\pvec{r}')\,\mathrm{d}^2r',
\end{split}
\end{align}
where $\delta$ is the 2D Dirac delta function and each vortex $v$ is associated with a flux $\Phi_v$ (typically $\Phi_v=n\Phi_0=nh/2e$, where $n$ is an integer, $\Phi_0$ is the magnetic flux quantum, $h$ is the Planck constant, and $e$ is the elementary charge). By solving Eq.~\ref{eq:applied_to_stream_vortices} to obtain $g(\vec{r})$, one can compute the supercurrent density in the film due to an applied field and flux trapped in both holes and vortices. For a simply-connected region $S$ containing a set of vortices $v$ each associated with a flux $\Phi_v$, the fluxoid is equal to $\Phi^f_S=\sum_{\mathrm{vortices}\,v}\Phi_v$. The numerical solution to Eq.~\ref{eq:applied_to_stream_vortices} is described at the end of Section~\ref{section:implementation}.

\subsection{Multi-layer structures}
\label{section:model:multilayer}

For structures with multiple films lying in different planes or layers, with layer $\ell$ lying in the plane $z=z_\ell$,
the stream functions and fields for all layers can be computed self-consistently using the following recipe:

\begin{enumerate}
    \item{
        Calculate the stream function $g_\ell(\vec{r})$ for each layer $\ell$ by solving Eq.~\ref{eq:applied_to_stream_vortices} given an applied field $H_{z,\,\mathrm{applied}}(\vec{r}, z_\ell)$.
    }
    \item{
        For each layer $\ell$, calculate the $z$-component of the field due to the currents in all other layers $m\neq\ell$ (encoded in the stream function $g_m(\vec{r})$)
        using Eq.~\ref{eq:field_from_kernel}.
    }
    \item{
        Re-solve Eq.~\ref{eq:applied_to_stream_vortices} taking the new applied field at each layer to be the original applied field plus the sum of screening fields from all other layers. This is accomplished via the substitution
        \begin{align}
        \begin{split}
            H_{z,\,\mathrm{applied}}(\vec{r}, z_\ell)
            &\to H_{z,\,\mathrm{applied}}(\vec{r}, z_\ell)\\
            &+ \sum_{m\neq\ell}
            \int_{F_m} Q_z(\vec{r},\pvec{r}')g_m(\pvec{r}')\,\mathrm{d}^2r',
            \label{eq:iterative}
        \end{split}
        \end{align}
        where $F_m$ is surface of all films in layer $m$ and $g_m$ is the stream function for layer $m$.
    }
    \item{
        Repeat steps 1-3 until the solution converges.
    }
\end{enumerate}

Convergence can be quantified by, for example, calculating the total magnetic flux through all films and holes in the model at the end of each iteration. In general, the more layers there are in a structure the more iterations are  required to reach a given level of convergence.

\section{Numerical Implementation}
\label{section:implementation}

In order to numerically solve Eq.~\ref{eq:field_from_kernel} and Eq.~\ref{eq:Heff_sub}, we have to discretize the films, holes, and the vacuum regions surrounding them. We use a triangular
(Delaunay) mesh, consisting of $p$ points (or vertices)
which together form $t$ triangles. Below we denote column vectors and matrices using bold font. $\mathbf{A}\mathbf{B}$
denotes matrix multiplication, with $(\mathbf{A}\mathbf{B})_{ij}=\sum_{k=1}^\ell A_{ik}B_{kj}$
($\ell$ being the number of columns in $\mathbf{A}$ and the number of rows in $\mathbf{B}$). Column vectors are treated as matrices with $\ell$ rows and 1 column. We denote element-wise multiplication with a lower dot, $(\mathbf{A}.\mathbf{B})_{ij}=A_{ij}B_{ij}$, and $\mathbf{A}^T$ denotes the transpose of matrix $\mathbf{A}$.

The discrete version of Eq.~\ref{eq:field_from_kernel} is
\begin{align}
\begin{split}
    \label{eq:field_from_kernel_num}
    \underbrace{\mathbf{h}_z}_\text{total field}
    &= \underbrace{\mathbf{h}_{z,\,\mathrm{applied}}}_\text{applied field}
    + \underbrace{(\mathbf{Q}.\mathbf{w}^T)\mathbf{g}}_\text{screening field}\\
    h_{z, i} &= h_{z,\,\mathrm{applied}, i} + \sum_j Q_{ij}w_jg_j,
\end{split}
\end{align}
where for clarity we show both the matrix version of Eq.~\ref{eq:field_from_kernel} (top line) and the equivalent discrete sum version (bottom line).

The $p\times p$ kernel matrix $\mathbf{Q}$ represents the kernel function $Q_z(\vec{r},\pvec{r}')$ for all points lying in the plane of the film, and the $p\times 1$ weight vector $\mathbf{w}$, which assigns an effective area to each vertex in the mesh, represents the differential element $\mathrm{d}^2r'$. Both $\mathbf{Q}$ and $\mathbf{w}$ are solely determined by the geometry of the mesh, so they only need to be computed once for a given device. $\mathbf{h}_z$, $\mathbf{h}_{z,\,\mathrm{applied}}$, and $\mathbf{g}$ are all $p\times 1$ vectors, with each row representing the value of the quantity at the
corresponding vertex in the mesh. The vector $\mathbf{w}$ is equal to the diagonal of the ``lumped mass matrix'' $\mathbf{M}$: $w_i=M_{ii} = \frac{1}{3}\sum_{t\in\mathcal{N}(i)}\mathrm{area}(t)$,
where $\mathcal{N}(i)$ is the set of triangles $t$ adjacent to vertex $i$. The kernel matrix $\mathbf{Q}$ is given by
\begin{equation}
    \label{eq:kernel_matrix}
    Q_{ij} = (\delta_{ij}-1)q_{ij}
    + \delta_{ij}\frac{1}{w_{j}}\left(C_i + \sum_{l\neq i}q_{il}w_{l}\right),
\end{equation}
where $q_{ij} = \left(4\pi|\vec{r}_i-\vec{r}_j|^3\right)^{-1}$
(which is $\lim_{\Delta z\to 0}Q_z(\vec{r},\pvec{r}')$ cf. Eq.~\ref{eq:kernels}),
and $\delta_{ij}$ is the Kronecker delta function. The diagonal terms involving the $p\times 1$ vector $\mathbf{C}$ are meant to work around the fact that $q_{ii}$ diverge (see Ref.~\cite{Brandt2005-wj} for more details), and $\mathbf{C}$ is given by
\begin{equation}
    \label{eq:C_vector}
    C_i = \frac{1}{4\pi}\sum_{p,q=\pm1}\sqrt{[\Delta x - p(x_i-\bar{x})]^{-2} + [\Delta y - q(y_i-\bar{y})]^{-2}},
\end{equation}
where $\Delta x=(x_\mathrm{max}-x_\mathrm{min})/2$ and $\Delta y=(y_\mathrm{max}-y_\mathrm{min})/2$ are half the side lengths of a rectangle bounding the modeled film and $(\bar{x}, \bar{y})$ are the coordinates of the center of the rectangle.

The matrix version of Eq.~\ref{eq:Heff_sub} is
\begin{equation}
    \label{eq:Heff_sub_num}
     \mathbf{h}_{z,\,\mathrm{applied}} - \sum_{\mathrm{holes}\, h}\mathbf{h}_{z,\,\mathrm{eff},\,h} = -(\mathbf{Q}.\mathbf{w}^T-\Lambda\mathbf{\nabla}^2)\mathbf{g},
\end{equation}
where we exclude points in the mesh lying outside of the superconducting film but keep points
inside holes in the film. $\mathbf{\nabla}^2$
is the Laplace operator, a $p\times p$ matrix defined such that $\mathbf{\nabla}^2\mathbf{f}$ computes the Laplacian $\nabla^2f(x,y)$ of a scalar field $f(x,y)$ defined on the mesh vertices (see~\ref{appendix:laplace}).

Eq.~\ref{eq:Heff_sub_num} is a matrix equation relating the applied field to the stream function
inside a superconducting film, which can efficiently be solved (e.g. by Cholesky or LU decomposition) for the unknown vector $\mathbf{g}$, the stream function inside the film. Since the stream function outside the film and inside holes in the film is already known, solving Eq.~\ref{eq:Heff_sub_num} gives us the stream function for the full mesh. Defining $\mathbf{K} = \left(\mathbf{Q}\cdot\mathbf{w}^T-\Lambda\mathbf{\nabla}^2\right)^{-1}$, we have
\begin{equation}
    \label{eq:full_stream}
    \mathbf{g} = \begin{cases}
        -\mathbf{K}
        \left(\mathbf{h}_{z,\,\mathrm{applied}} - \sum_{\mathrm{holes}\,h}\mathbf{h}_{z,\,\mathrm{eff},\,h}\right)
            & \text{inside the film}\\
        I_{\mathrm{circ},\,h}
            & \text{inside hole }h\\
        0
            & \text{elsewhere}
    \end{cases}
\end{equation}

If there is a vortex containing flux $\Phi_j$ located in a film at position $\vec{r}_j$ indexed as mesh vertex $j$, then for each position $\vec{r}_i$ within that film, we add to the stream function $g_i$ the quantity $\mu_0^{-1}\Phi_jK_{ij} / w_{j}$, where $K_{ij}$ is an element of the inverse matrix defined above, and $w_{j}$ is an element of the weight matrix which assigns an effective area to the mesh vertex at which the vortex is located. This process amounts to numerically inverting Eq.~\ref{eq:applied_to_stream_vortices} as described in Ref.~\cite{Brandt2005-wj}.

Once the stream function $\mathbf{g}$ is known for the full mesh,
the supercurrent flowing in the film can be computed from Eq.~\ref{eq:stream}, the $z$-component of the total field in the plane of the film can be computed
from Eq.~\ref{eq:field_from_kernel_num}, and the full vector magnetic field $\vec{H}(x, y, z)$
at any point in space can be computed from Eqs.~\ref{eq:field_from_kernel} and ~\ref{eq:kernels}. Multi-layer structures are solved iteratively as described in Section \ref{section:model:multilayer}.

\section{Package Overview}
\label{section:overview}

In this section we give a high-level overview of the \SuperScreen package. Further details can be found in the online documentation~\cite{superscreen-rtd}. The specific version of the package corresponding to this manuscript is \texttt{v0.5.0}.

\subsection{Development Details}
\label{section:overview:development}

At the time of writing, \SuperScreen requires Python version 3.7--3.9. The package is located in a public repository on GitHub~\cite{superscreen, BishopVanHorn2022-bd}, and a suite of unit tests is run automatically via the GitHub Actions continuous integration (CI) tool whenever a change or proposed change (Pull Request) is made to the \inline{main} branch of the repository. At the time of writing, the test suite is executed using Python versions 3.7 through 3.9, and the test coverage is $>95\%$. Any changes to the \inline{main} branch of the repository also trigger an automatic re-build of the online documentation~\cite{superscreen-rtd}. Stable versions of the package are tagged on GitHub and uploaded to PyPI, the Python Package Index. The source code and documentation are provided under the MIT License.\footnote{\href{https://opensource.org/licenses/MIT}{https://opensource.org/licenses/MIT}}

\SuperScreen has several important dependencies beyond the Python standard library: \inline{numpy}~\cite{Harris2020-xv} and \inline{scipy}~\cite{Virtanen2020-zz} for numerics, \inline{matplotlib}~\cite{Hunter2007-il} for visualization, \inline{pint}~\cite{Grecco} for handling physical units, \inline{shapely}~\cite{shapely} for creating and manipulating device geometries, \inline{meshpy}~\cite{Klockner, Shewchuk, Shewchuk1996-va} and \inline{optimesh}~\cite{Schlomer2021-ua} for mesh generation, and \inline{Ray}~\cite{Moritz2018-mt,ray-docs} for parallel processing with shared memory (see~\ref{section:parallel}).

\subsection{Devices}
\label{section:overview:device}

\begin{table*}
\centering
\begin{tabular}{|l|l|l|}
\hline
\SuperScreen                             & Set-theoretic  & Boolean   \\ \hline \hline
\inline{polygonA.union(polygonB)}        & $A\cup B$      & \inline{OR}      \\
\inline{polygonA.intersection(polygonB)} & $A\cap B$      & \inline{AND}     \\
\inline{polygonA.difference(polygonB)}   & $A\setminus B$ & \inline{AND NOT} \\
\inline{polygonA.difference(polygonB, symmetric=True)} & $(A\setminus B)\cup(B\setminus A)$ & \inline{XOR} \\ \hline
\end{tabular}
\caption{Methods for combining \inline{superscreen.Polygon} objects, along with their corresponding set-theoretic and boolean logic operations.}
\label{table:polygon}
\end{table*}

Information about the geometry and penetration depth of a superconducting structure is described by an instance of the \inline{superscreen.Device} class. A \inline{Device} is made up of one or more superconducting layers, each represented by an instance of  \inline{superscreen.Layer}. Each layer sits in a specified plane parallel to the $x-y$ plane and has its own effective penetration depth $\Lambda$. Alternatively, the effective penetration depth $\Lambda$ can be defined in terms of a layer's London penetration depth $\lambda$ and its thickness $d$: $\Lambda=\lambda^2/d$.

Each layer can contain one or more superconducting \inline{films} which may have one or more \inline{holes} in them. Films and holes are represented by instances of the \inline{superscreen.Polygon} class. All polygons in a device must be simply-connected; a hole in a film is modeled as one \inline{Polygon} instance whose coordinates all lie within the \inline{Polygon} representing the film. \inline{Polygons} can be constructed and combined using set-theoretic operations. Table~\ref{table:polygon} shows the four methods available for combining a \inline{supercreen.Polygon} instance \inline{polygonA}, whose vertices lie in set $A$, with a polygon \inline{polygonB}, whose vertices lie in set $B$. Note that \inline{polygonB} can be a \inline{superscreen.Polygon}, an $n\times 2$ \inline{numpy} array of vertex coordinates, or a \inline{LineString}, \inline{LinearRing}, or \inline{Polygon} from the \inline{shapely} package~\cite{shapely}.

In addition to superconducting films and holes, one may define ``abstract regions,'' which are polygons that do not necessarily correspond to a physical feature in the structure, but will still be meshed. Abstract regions can be used to define a ``bounding box" around a structure to be modeled, or to locally increase the density of the computational mesh in a given region. The \inline{superscreen.geometry} module provides functions for generating the underlying polygon vertices for simple shapes (ellipses and rectangles), which can be combined as described above to create more complicated geometries.

Once the layers, films, holes, and abstract regions have been defined, one can generate the computational mesh by calling \inline{Device.make_mesh()}. The region that is meshed is defined by the convex hull of the union of all polygons in the device. Mesh generation is a two step process. First, an initial Delaunay mesh is created using \inline{meshpy}~\cite{Klockner}, which is a Python interface to Triangle~\cite{Shewchuk1996-va, Shewchuk}, a fast compiled 2D mesh generation tool. Second (and optionally), the mesh is further optimized using \inline{optimesh}~\cite{Schlomer2021-ua}. The goal of the mesh optimization step is to improve the ``quality'' of each triangular element in the mesh, where quality measures how close a triangle is to equilateral (quality $\leq1$, with equality for equilateral triangles). In practice, the more \inline{optimesh} steps are performed, the more uniform in size and spatial density the triangles in the mesh become. The local density of triangles in the mesh is determined by the density of vertices in the device's polygons and the total number of triangles. Regions where there are many polygon vertices will be meshed more densely than regions with few polygon vertices. If no \inline{optimesh} optimization is performed, then every polygon vertex is guaranteed to be a mesh vertex. See Code Block~\ref{code:device} for a demonstration of the process of creating a \inline{Device}, and Figure~\ref{fig:ring-with-slit}({\bf a}) to view the resulting geometry and mesh. After the mesh has been generated, the geometry-dependent matrices and vectors described in Section~\ref{section:implementation} are computed and one can begin solving models.

\begin{code}
\begin{minted}[fontsize=\small]{python}
import superscreen as sc
from superscreen.geometry import circle, box

# Define the device geometry.
length_units = "um"
ro = 3  # outer radius
ri = 1  # inner radius
slit_width = 0.25
Lambda = 1  # effective penetration depth
# circle() and box() generate arrays of polygon (x, y) coordinates.
ring = circle(ro)
hole = circle(ri)
slit = box(slit_width, 1.5 * (ro - ri), center=(0, -(ro + ri) / 2))
# Define the Polygon representing the superconductor.
layer = sc.Layer("base", Lambda=Lambda)
film = sc.Polygon.from_difference(
    [ring, slit, hole], name="ring_with_slit", layer="base"
)
bounding_box = sc.Polygon("bounding_box", layer="base", points=circle(1.2 * ro))
# Create a Device and generate and plot the computational mesh.
device = sc.Device(
    film.name,
    layers=[layer],
    films=[film],
    abstract_regions=[bounding_box],
    length_units=length_units,
)
device.make_mesh(min_points=3500)
device.plot(mesh=True)
# Calculate the device's response to a uniform applied field.
applied_field = sc.sources.ConstantField(10)
solution = sc.solve(device, applied_field=applied_field, field_units="mT")[-1]
# Visualize the solution.
# Plot the current density evaluated at each layer in the Device.
solution.plot_currents()
# Plot the magnetic field evaluated at each layer in the Device.
solution.plot_fields()
# Plot the field evaluated at any points in space.
solution.plot_field_at_positions(device.points, zs=0.5)
\end{minted}
\captionof{listing}{The typical workflow for a \SuperScreen simulation: 1) Define the device geometry and materials properties, 2) generate the computational mesh, 3) solve the model for a given applied field and/or trapped flux, and 4) visualize the results.}
\label{code:device}
\end{code}

\subsection{Solvers}
\label{section:overview:solvers}
\begin{figure*}
    \centering
    \includegraphics[width=\textwidth]{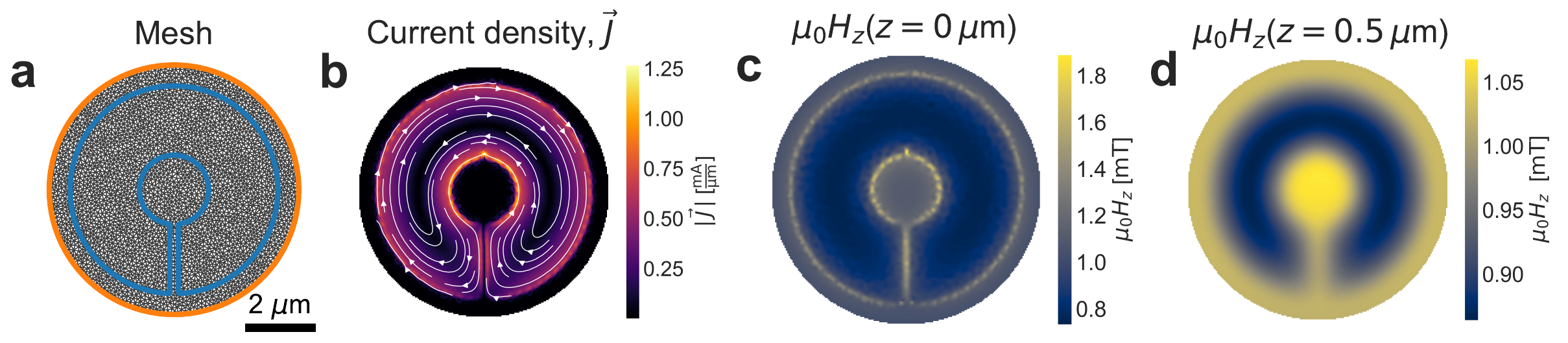}
    \caption{The output of Code Block~\ref{code:device}: Meissner screening of a uniform $1\,\mathrm{mT}$ out-of-plane field by a ring with inner diameter $1\,\um$, outer diameter $3\,\um$, and effective penetration depth $\Lambda=1\,\um$, interrupted by a slit of width $0.25\,\um$. ({\bf a}) Plot of the boundary of the ring (blue), circular bounding box (orange), and the computational mesh (gray), generated with \inline{Device.plot()}. ({\bf b}) The current density $\vec{J}$ in the ring, generated with \inline{Solution.plot_currents()}. ({\bf c}) The $z$-component of the magnetic field $\mu_0H_z$ evaluated at the plane of the ring, generated with \inline{Solution.plot_fields()}. ({\bf d}) The $z$-component of the magnetic field evaluated $z=0.5\,\um$ above the ring (generated using \inline{Solution.plot_field_at_positions()}).}
    \label{fig:ring-with-slit}
\end{figure*}

A \SuperScreen model consists of 1) a \inline{Device} with a mesh, 2) a function or \inline{Parameter} that defines the applied magnetic field as a function of position $H_{z,\,\mathrm{applied}}(x, y, z)$, 3) a value for the current circulating around each hole in the device due to trapped flux, and 4) a collection of vortices $v$ located at positions $\vec{r}_v$ and carrying flux $\Phi_v$. These items serve as the inputs to \SuperScreen's main solver function, \inline{superscreen.solve()}, which implements the calculation outlined in Section~\ref{section:implementation}. When simulating a device with more than one layer, one can specify the number of times to implement the iterative calculation described in Section~\ref{section:model:multilayer} in order to solve for the response of all layers self-consistently. One can also skip the iterative portion of the calculation entirely and only solve for the response of each layer to the applied field, assuming no interaction between layers. The \inline{device.solve_dtype} attribute determines the \inline{numpy} floating point data type used by \inline{solve()}. The default data type is \inline{float64} (64-bit double-precision float, equivalent to Python's \inline{float} type), but one can, for instance, set \inline{device.solve_dtype = "float32"} to use 32-bit single-precision floats in order to save memory.

The output of \inline{superscreen.solve()} is a \inline{list} of \inline{superscreen.Solution} objects, with a length of 1 plus the number of iterations used for the iterative portion of the calculation. A \inline{Solution} encapsulates all of the information about a solved model: the \inline{Device}, applied field, circulating currents, vortices, and calculated stream functions and magnetic fields for all layers in the device. A \inline{Solution} also has methods for processing the simulation results, including:
\begin{itemize}
    \item{
    \inline{Solution.grid_data()}: Interpolates the calculated stream functions $g(x, y)$, magnetic fields $\mu_0H_z(x, y)$, or current densities $\vec{J}(x, y)$, for each layer from the triangular mesh to a rectangular grid.
    }
    \item{
    \inline{Solution.field_at_position()}: Calculates the vector magnetic field at any point(s) in space due the applied field and the currents flowing the in the device using Eqs.~\ref{eq:field_from_kernel} and \ref{eq:kernels}.
    }
    \item{
    \inline{Solution.interp_current_density()}: Evaluates the 2D current density $\vec{J}(x, y)$ in each layer at arbitrary $(x, y)$ coordinates via interpolation.
    }
    \item{
    \inline{Solution.polygon_flux()}: Calculates the total flux through each polygon in the device.
    }
    \item{
    \inline{Solution.polygon_fluxoid()}: Calculates the fluxoid for a specified polygonal region in the device. See Section~\ref{section:examples:fluxoid} for more details.
    }
\end{itemize}
\inline{Solutions} also have several visualization methods built in (see Code Block~\ref{code:device}, Figure~\ref{fig:ring-with-slit}, and Section~\ref{section:overview:visualization}).

One may wish to solve many models involving the same device while varying other aspects of the model, for example sweeping the applied field, circulating currents, vortex properties, or some parameter of one or more layers in the device. Fortunately, the mesh, Laplace operator, kernel matrix, etc. (described in Section~\ref{section:implementation}) depend only on the geometry of the device parallel to the $x-y$ plane. This means that the same mesh and matrices can be re-used for models with different applied fields, circulating currents, vortex properties, layer $z$-positions, and penetration depths.

The \inline{superscreen.solve_many()} function manages the setup and execution of such a sweep. One can provide a sequence of \inline{Parameter} objects defining different applied fields and/or a sequence of circulating current values over which to sweep and/or a ``layer updater" function that modifies each layer in the device according to some set of keyword arguments, which can also be swept. The latter option can be used to sweep layer heights or penetration depths. Given these inputs, \inline{superscreen.solve_many()} will generate and solve all of the corresponding models. The models can either be solved in series in a single Python process (the default), or in parallel in multiple Python processes running across multiple CPUs, or even across multiple nodes in a cluster (see~\ref{section:parallel}).

\subsection{Visualization}
\label{section:overview:visualization}

\SuperScreen offers several functions for visualizing the results of simulations (which are also aliased as methods on \inline{superscreen.Solution}):

\begin{itemize}
    \item{
    \inline{superscreen.plot_streams()}: Given a \inline{Solution}, plots the stream function $g(x, y)$ for one or more layers in the device.
    }
    \item{
    \inline{superscreen.plot_currents()}: Given a \inline{Solution}, plots the current density $\vec{J}(x, y)$ for one or more layers in the device.
    }
    \item{
    \inline{superscreen.plot_fields()}: Given a \inline{Solution}, plots the total field $H_z(x, y)$ or the screening field $H_z(x, y) - H_{z,\,\mathrm{applied}}(x, y)$ for one or more layers in the device.
    }
    \item{
    \inline{superscreen.plot_field_at_positions()}: Given a \inline{Solution}, plots the total field $\vec{H}(x, y, z)$ or $H_z(x, y, z)$ at an arbitrary set of positions $(x, y, z)$.
    }
\end{itemize}

See Code Block~\ref{code:device} and  Figure~\ref{fig:ring-with-slit} for an example of the usage and output of \inline{plot_fields()} and \inline{plot_currents()}.

\subsection{Comparison \& Persistence}
\label{section:overview:persistence}

\inline{Parameters}, \inline{Layers}, \inline{Polygons}, \inline{Devices}, and \inline{Solutions} all implement the equality operator, \inline{==}. Two \inline{Parameters} are considered equal if the Python bytecode of their underlying functions is the same and their keyword arguments are the same. Two \inline{Layers} are equal if their name, penetration depth, thickness, and vertical position are all equal. Two \inline{Polygons} are equal if they are in the same layer and their name and polygon vertices are equal. Two \inline{Devices} are equal if their name, layers, films, holes, and abstract regions are all equal. Two \inline{Solutions} are equal if their device, applied field, circulating currents, list of trapped vortices, timestamp (time at which the solution was created), and all stream function and magnetic field arrays are equal. Two \inline{Solutions} created at different times can also be compared using the  \inline{solution.equals()} method.

Instances of \inline{superscreen.Device} and \inline{superscreen.Solution} can be saved to and loaded from disk using their respective \inline{to_file()} and \inline{from_file()} methods, making it straightforward to store and share models and simulation results. \inline{Layers}, \inline{Polygons}, and all metadata are serialized to JSON, a widely-used, human-readable plain text format. Functions and \inline{Parameters}, such as those that compute the applied field or penetration depth, are serialized in binary form using the \inline{dill} package~\cite{McKerns}. \inline{Numpy} arrays, such as the mesh itself and the computed stream functions and fields, are saved in the \inline{numpy} \inline{npz} file format. A \inline{list} of \inline{Solutions}, such as that returned by \inline{superscreen.solve()} can be saved/loaded all at once using \inline{save_solutions()} and \inline{load_solutions()}.

\section{Examples}
\label{section:examples}

\subsection{Calculating the fluxoid}
\label{section:examples:fluxoid}

\SuperScreen allows one to calculate the fluxoid $\Phi^f_S$ for any polygonal region $S$ whose boundary $\partial S$ lies completely within a superconducting film using the method \inline{Solution.polygon_fluxoid()} (see Section~\ref{section:model:fluxoid}, Eq.~\ref{eq:fluxoid}). The ``flux part''  $\int_S\mu_0H_z(\vec{r})\,\mathrm{d}^2r$ is calculated using \inline{Solution.polygon_flux()}, which computes the flux through a polygon representing a region $S$ as $\Phi_S=\sum_{i\in S} \mu_0H_{z,i}w_i$, where $H_{z,i}$ is the magnetic field at vertex $i$ (recall that $w_i$ assigns an effective area to mesh vertex $i$). The ``supercurrent part'' $\oint_S\mu_0\Lambda\vec{J}(\vec{r})\cdot\mathrm{d}\vec{r}$ is calculated by evaluating the vector current density $\vec{J}$ at each point in the path $\partial S$ using \inline{Solution.interp_current_density()}, then computing the line integral along the path using the trapezoid rule. The sum of these two terms gives the fluxoid $\Phi^f_S$.

\begin{figure}[t]
    \centering
    \includegraphics[width=\linewidth]{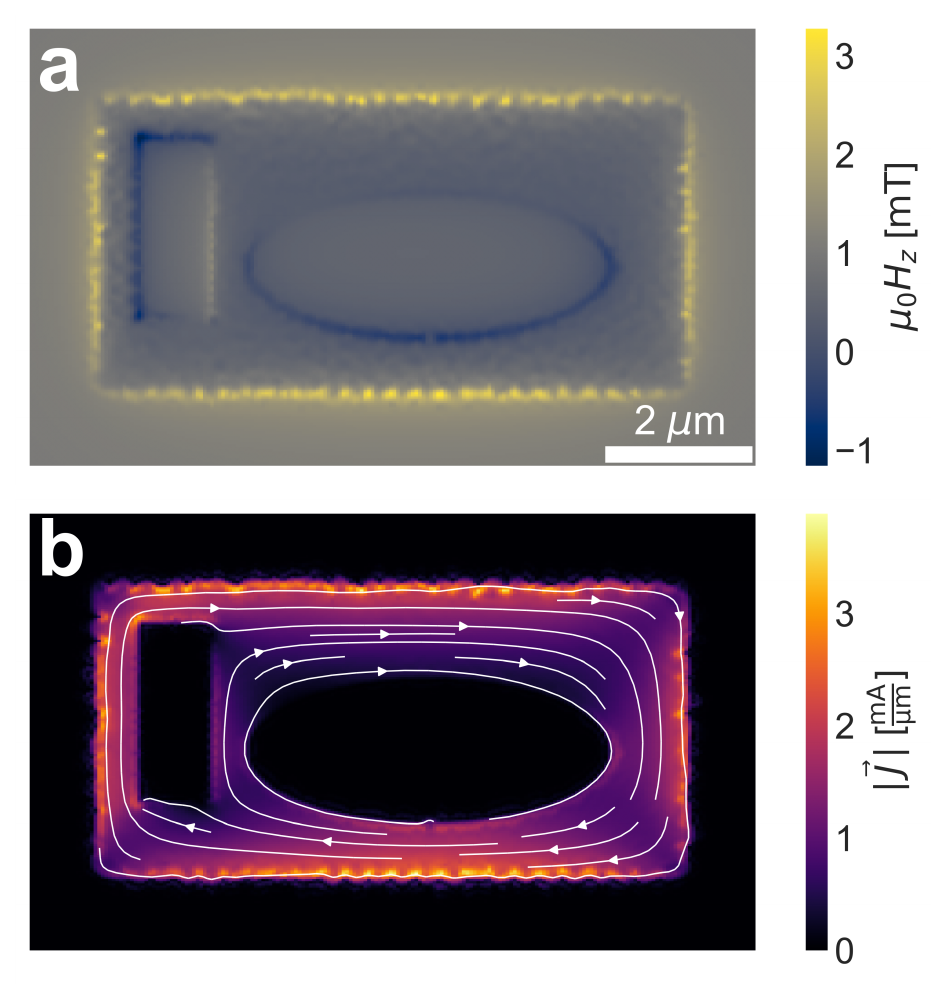}
    \caption{({\bf a}) Magnetic field and ({\bf b}) current density distributions generated by Code Block~\ref{code:fluxoid} for a rectangular superconducting film with $\Lambda=0.25\,\um$, which has one rectangular and one elliptical hole. The film is subject to a uniform applied out-of-plane field of 1 mT, and both holes are set to be in the $n=0$ fluxoid (i.e. Meissner) state. The resulting ciruclating currents are $I_\text{rectangle} = -1.071$ mA and $I_\text{ellipse} = -1.589$ mA, and the residual fluxoid for each hole is smaller than $10^{-7}\,\Phi_0$. These results were computed using a mesh with approximately 5,000 vertices and 10,000 triangles.}
    \label{fig:fluxoid}
\end{figure}

While the 2D London model doesn't ``know'' about fluxoid quantization, in the sense that the quantization condition $\Phi^f_S=n\Phi_0$ is not automatically satisfied by solutions to Eq.~\ref{eq:london} for multiply-connected films, we can nevertheless calculate current and field distributions for different fluxoid states in multiply-connected superconductors by adjusting the currents circulating around each hole to realize a prescribed set of fluxoid values. For a structure with $N_h$ holes, we can specify $N_h$ fluxoids $\Phi^f_h$ and find the circulating currents $I_h$ by minimizing the deviation of each fluxoid from its desired value. This calculation is implemented in the \inline{superscreen.find_fluxoid_solution()} function. For $N_h=1$, it is treated as a root-finding problem, which can be solved with typically only three calls to \inline{superscreen.solve()}. For $N_h>1$, it is a least-squares minimization problem with $N_h$ free parameters. Code Block~\ref{code:fluxoid} demonstrates how to model a device with one or more holes, each in the $n=0$ fluxoid state, subject to a uniform applied field, and Figure~\ref{fig:fluxoid} shows the field and current distributions for a rectangular superconducting film with $\Lambda=0.25\,\um$, which has one rectangular and one elliptical hole. The least-squares minimization for the model shown in Figure~\ref{fig:fluxoid}, with $N_h=2$, required 18 total calls to \inline{superscreen.solve()}. It is important to note that while \SuperScreen can calculate the field and current distributions for a given fluxoid state, the model does not capture transitions between fluxoid states.

\begin{code-onecol}
\begin{minted}[fontsize=\small]{python}
import superscreen as sc
# Assume that we have already created a
# superscreen.Device with one or more
# holes, and generated the mesh.
# Specify the desired fluxoid for each
# hole in the device:
fluxoids = {
    hole_name: 0
    for hole_name in device.holes
}
applied_field = sc.sources.ConstantField(1)
field_units = "mT"

result = sc.find_fluxoid_solution(
    device=device,
    fluxoids=fluxoids,
    applied_field=applied_field,
    field_units=field_units,
)
solution, opt_result = result
\end{minted}
\captionof{listing}{Calculating the field and current distributions for fluxoid states in a multiply-connected superconducting film. Given a \inline{Device} with $N_h\geq 1$ holes and a desired fluxoid $\Phi^f_h$ for each hole, \inline{superscreen.find_fluxoid_solution()} optimizes the current $I_h$ circulating around each hole to realize the desired fluxoid state. The function returns a \inline{tuple} of length 2, the first element being the final optimized \inline{superscreen.Solution} and the second element being an instance of either \inline{scipy.optimize.RootResults} (if $N_h=1$) or \inline{scipy.optimize.OptimizeResult} (if $N_h>1$), which contains information about the optimization that was performed. See Figure~\ref{fig:fluxoid} for an example of the results for a film with two holes, both in the $n=0$ fluxoid state.}
\label{code:fluxoid}
\end{code-onecol}

\subsection{Pearl vortices in thin films}
\label{section:examples:pearl-vortices}
Vortices trapped in 2D superconductors ($d\ll\lambda$, where $d$ is the film thickness and $\lambda$ is the London penetration depth), i.e. ``Pearl vortices,'' are associated with different current and magnetic field distributions than Abrikosov vortices trapped in bulk type-II superconductors~\cite{Pearl1964-cl}. The 2D Fourier transform $\tilde{H}_z(\vec{k}, z)$ of the out-of-plane component of the magnetic field $H_z(\vec{r}, z)$ from a Pearl vortex located at the origin $x=y=z=0$ is given by
\begin{equation}
    \tilde{H}_z(\vec{k}, z)=\mathcal{F}\{H_z(\vec{r}, z)\}=\frac{1}{\mu_0}\frac{\Phi_0e^{-|\vec{k}|z}}{1+2\Lambda|\vec{k}|},
    \label{eq:pearl}
\end{equation}
where $\mathcal{F}\{\cdot\}$ is the 2D Fourier transform, $\vec{k}=(k_x, k_y)$ are in-plane spatial frequencies, $z$ is the out-of-plane position at which the field is evaluated, and $2\Lambda = 2\lambda^2 / d$ is the Pearl length~\cite{Pearl1964-cl, Tafuri2004-ap}. The real-space magnetic field distribution near a Pearl vortex can be calculated by taking the inverse Fourier transform of Eq.~\ref{eq:pearl}: $H_z(\vec{r}, z)=\mathcal{F}^{-1}\{\tilde{H}_z(\vec{k}, z)\}$.

\begin{figure*}[h!]
    \centering
    \includegraphics[width=\textwidth]{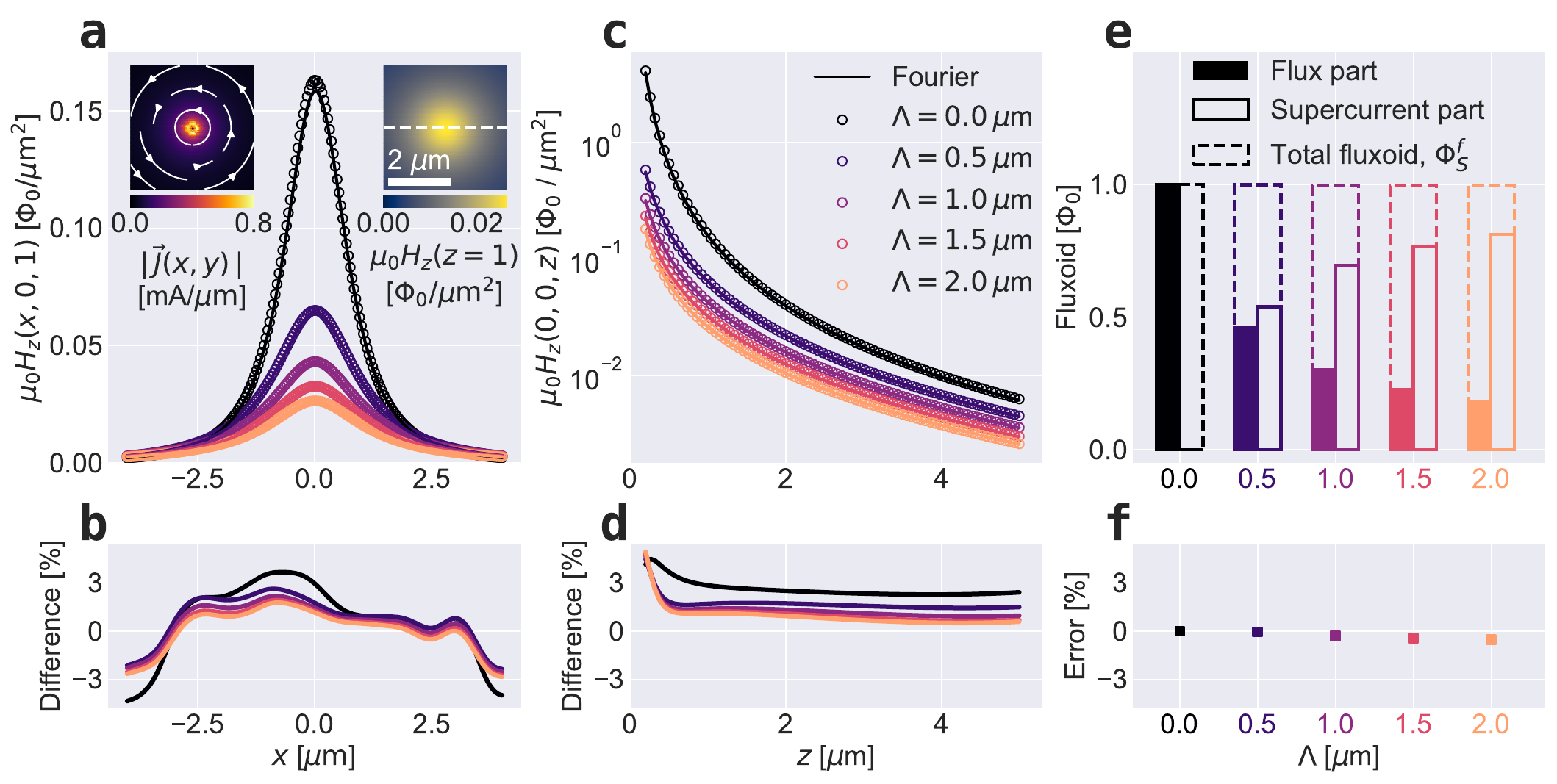}
    \caption{Vortex field profiles calculated using \SuperScreen agree with the analytical solution (Eq.~\ref{eq:pearl}) to within a few percent. Here, we model a vortex trapped at the center of a square superconducting film lying in the $x-y$ plane with side length $20\,\um$ as a function of the film's effective penetration depth, $\Lambda$. ({\bf a}) Cross section along the $x$-axis of the out-of-plane magnetic field $\mu_0H_z$ from the vortex, evaluated at a vertical distance $z=1\,\um$ above the film. The left inset shows the current density in the plane for film for $\Lambda=2\,\um$, and the right inset shows the corresponding magnetic field evaluated at $z=1\,\um$, with a dashed white line indicating the cross-section axis. ({\bf b}) Percentage difference between the $\mu_0H_z$ calculated with \SuperScreen and the Fourier transform method for the $x$-axis cut shown in ({\bf a}). ({\bf c}) Cross-section along the $z$-axis of $\mu_0H_z$ directly above the center of the vortex (logarithmic $y$ axis scale). ({\bf d}) Percentage difference between the $\mu_0H_z$ calculated with \SuperScreen and the Fourier transform method for the $z$-axis cut shown in ({\bf c}). In ({\bf a}) and ({\bf c}), the results from \SuperScreen are shown as open circles and the results from the Fourier transform method are shown as solid lines. ({\bf e}) The fluxoid $\Phi_S^f$ for a circular region in the film with radius $r=1\,\um$ centered on the vortex core. As $\Lambda$ increases, so to does the supercurrent contribution to the fluxoid. ({\bf f}) Error in the total simulated fluxoid relative to $\Phi_0$: error = $(\Phi_S^f(\Lambda) - \Phi_0) / \Phi_0$.}
    \label{fig:pearl}
\end{figure*}

To include vortices in a \SuperScreen model, one can simply input a \inline{list} of \inline{superscreen.Vortex} objects when calling \inline{superscreen.solve()}. A \inline{Vortex} object specifies the $x,y$ position for the vortex core, the name of the superconducting layer in which the vortex is pinned, and the number of flux quanta $\Phi_0$ contained in the vortex (which is 1 by default). The field distributions generated by \SuperScreen in the presence of vortices as described in Sections~\ref{section:model:vortices} and~\ref{section:implementation} agree to within a few percent with the distributions obtained using this Fourier transform method, as demonstrated in Figure~\ref{fig:pearl}. Figure~\ref{fig:pearl}({\bf e}) shows the fluxoid for a circular region $S$ with radius $r=1\,\um$ enclosing a Pearl vortex trapped in a film as a function of the film's effective penetration depth, $\Lambda$. When $\Lambda=0$, the screening currents decay very quickly away from the center of the vortex, so the ``supercurrent part'' of the $\Phi^f_S$ vanishes. With increasing $\Lambda$, the ``supercurrent part'' accounts for an increasing fraction of the total fluxoid. See Ref.~\cite{Brandt2005-wj} for a method to compute the self-energy and interaction energies of vortices in thin films.

\subsection{Calculating inductance}
\label{section:examples:inductance}

\begin{figure}
    \centering
    \includegraphics[width=\linewidth]{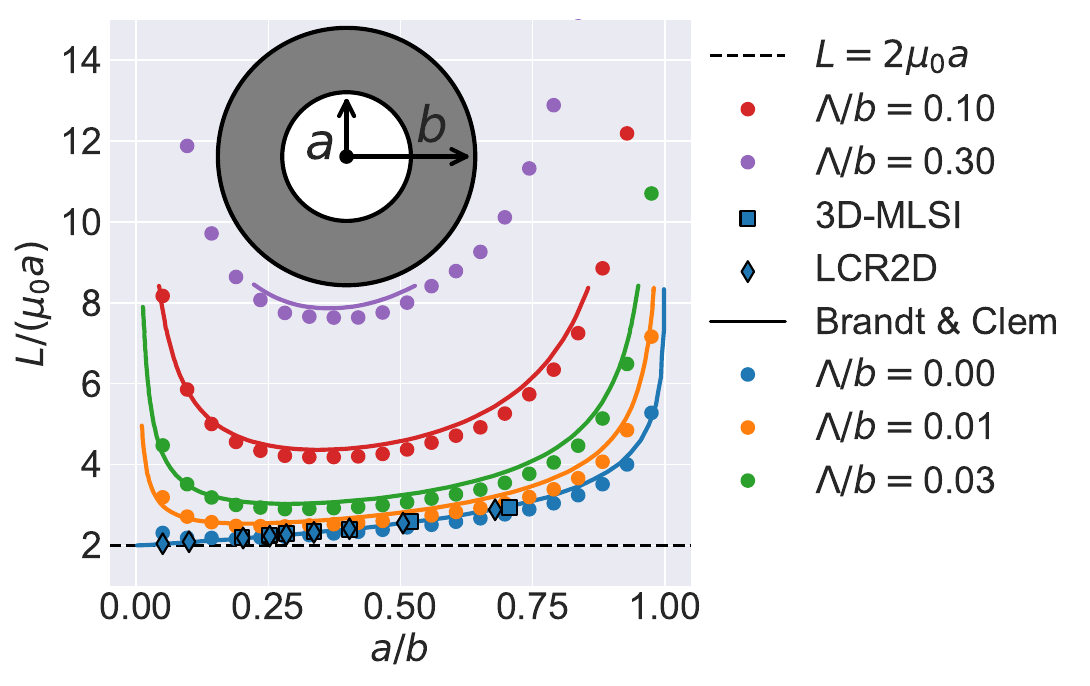}
    \caption{Self-inductance $L$ of a circular ring with inner radius $a$ and outer radius $b$ (see inset), as a function of the ratio $a / b$ and the ring's effective penetration depth $\Lambda$. Filled circles indicate results from \SuperScreen, solid lines show numerical results from Figure 2 of Ref.~\cite{Brandt2004-ew}, blue squares and diamonds show numerical results from Figure 1 of Ref.~\cite{Khapaev1997-kw}, and the dashed line indicates the analytical solution, $L = 2\mu_0 a$, for $\Lambda=0$ in the limit $a/b\to 0$\cite{Ketchen2012-mb, Babaei_Brojeny2003-la}. The \SuperScreen results were generated using a mesh with approximately 4,000 vertices and 8,000 triangles.}
    \label{fig:inductance}
\end{figure}

As shown in Ref.~\cite{Brandt2005-wj}, the mutual inductance $M_{ij}$ between holes $i$ and $j$ in a superconducting structure is given by
\begin{equation}
    M_{ij}=\frac{\Phi^f_{S_i}}{I_j},
\end{equation}
where $\Phi^f_{S_i}$ is the fluxoid for a region $S_i$ containing the hole $i$, and $I_j$ is the current circulating around hole $j$. The mutual inductance values for a set of holes form a mutual inductance matrix. The diagonals of the mutual inductance matrix are the hole self-inductances ($M_{ii}=L_i$, the self-inductance of hole $i$), and the matrix is symmetric ($M_{ij}=M_{ji}$) due to the reciprocity theorem. In this context, the flux and supercurrent parts of the fluxoid correspond to the geometric and kinetic inductance respectively~\cite{Brandt2004-ew}. If the penetration depth of the film containing hole $i$ is $\Lambda = 0$, then no field penetrates the film, the fluxoid $\Phi^f_{S_i}$ is equal to the flux through hole $i$, and the total inductance is equal to the geometric inductance. For a device with $N_h$ holes, the $N_h\times N_h$ mutual inductance matrix $\mathbf{M}$ can be computed using the $\inline{Device.mutual_inductance_matrix()}$ method. For example, the mutual inductance matrix for the device shown in Figure~\ref{fig:fluxoid} is:
\begin{equation*}
    \mathbf{M} = 
    \begin{pmatrix}
    	10.319 & -1.536 \\
        -1.527 & 7.130
    \end{pmatrix}
    \,\mathrm{pH},
\end{equation*}
where the matrix is indexed as $$\begin{pmatrix}\text{ellipse}\\\text{rectangle}\end{pmatrix},$$
and the ``fractional asymmetry'' of $\mathbf{M}$ is $|M_{01} - M_{10}| / \min(|M_{01}|, |M_{10}|)\approx 0.6\%$.

The self-inductance $L$ of a 2D circular ring with inner radius $a$ and outer radius $b$ (see inset of Figure~\ref{fig:inductance}) has been used as an informal benchmark for superconducting inductance calculations. For a ring with effective penetration depth $\Lambda = 0$, it has been shown analytically that $L\to 2\mu_0 a$ in the limit $a/b\to 0$~\cite{Ketchen2012-mb, Babaei_Brojeny2003-la}. Khapaev calculated the inductance for $\Lambda = 0$ as a function of $a/b$~\cite{Khapaev1997-kw}, and Brandt and Clem calculated the inductance as a function of both $\Lambda$ and $a / b$~\cite{Brandt2004-ew}. Figure~\ref{fig:inductance} shows a comparison between these previous numerical results and the results from \SuperScreen. Note that the models used by Brandt and Clem (solid lines) and by the LCR2D software (blue diamonds) require a circularly symmetric superconducting film, whereas \SuperScreen (filled circles) and 3D-MLSI~\cite{Khapaev1997-kw} (blue squares) support arbitrary 2D geometry. 

\subsection{Application: scanning SQUID susceptometry}
\label{section:examples:scanning-squid}

\begin{figure*}[t]
    \centering
    \includegraphics[width=\textwidth]{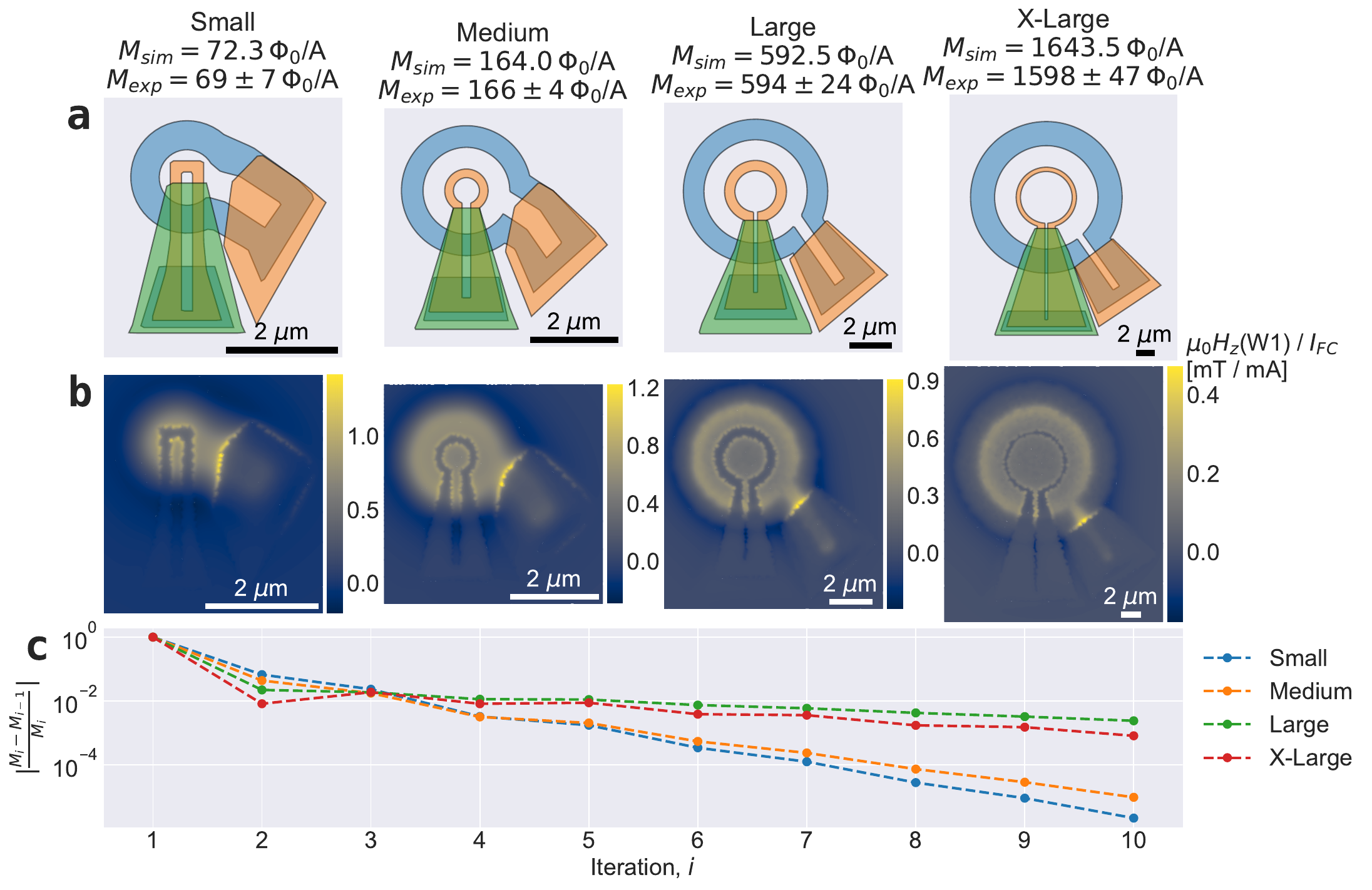}
    \caption{Calculating SQUID susceptometer mutual inductance, defined as the fluxoid induced in the SQUID pickup loop per unit current flowing in the field coil: $M_\mathrm{PL-FC}=\Phi^f_\mathrm{PL} / I_\mathrm{FC}$. Top row ({\bf a}): Schematics of \SuperScreen models for the field coil and pickup loop region of four sizes of scanning SQUID susceptometer (generated using \inline{Device.draw()}) with pickup loop inner radii ranging from $0.1\,\um$ (``Small'') to $3\,\um$ (``X-Large''). Middle row ({\bf b}): Simulated magnetic field $\mu_0H_z$ evaluated at the plane of layer W1, which contains the pickup loop, normalized by the current $I_\mathrm{FC}$ flowing in the field coil (which is located in layer BE). Bottom row ({\bf c}): Convergence of the four models, defined as the fractional change in mutual inductance between subsequent iterations $i$. For all four models, the simulated mutual inductance, $M_\mathrm{sim}$, falls within the range of mutual inductance values measured in real devices, $M_\mathrm{exp}$, which were reported in Table 1 of Ref.~\cite{Kirtley2016-zz}. The mutual inductance values are shown in units of $\Phi_0 / \mathrm{A}$, where $1\,\Phi_0 / \mathrm{A}\approx2.068\times10^{-3}\,\mathrm{pH}$. The meshes for all four models consisted of approximately 6,000 vertices and 12,000 triangles. The two smaller models converge more quickly than the two larger models for reasons discussed in \ref{appendix:numerics}. Note that for the ``X-Large'' model, we set the thicknesses of layers I1 and I2 both to 400 nm (instead of the nominal values of 150 nm and 130 nm respectively), to ensure convergence (see \ref{appendix:numerics} and Figure~\ref{fig:xlarge-dz-dr}).}
    \label{fig:squid-mutuals}
\end{figure*}

As an example application, we consider scanning SQUID microscopy, a technique in which a superconducting sensor is used to study superconducting or magnetic samples on micron length scales~\cite{Kirtley2016-zz}. In scanning SQUID susceptometry, the magnetic susceptibility of a sample is measured by bringing the sample close to a pair of superconducting loops~\cite{Gardner2001-gr, Huber2008-il, Kirtley_Kalisky_2012}. The first loop, called the ``pickup loop" (PL) is attached to a SQUID circuit that sensitively measures the magnetic flux threading the loop. The second loop, called the ``field coil" (FC), carries a known current $I_\mathrm{FC}$ and applies a known magnetic field to both the pickup loop and the sample. A superconducting sample will screen the magnetic field from the field coil, modifying the amount of flux threading the pickup loop and reducing the mutual inductance $M_\mathrm{PL-FC}$. The magnitude of this reduction in $M_\mathrm{PL-FC}$ is a measure of the sample's penetration depth and therefore its superfluid density.

\begin{figure*}
    \centering
    \includegraphics[width=\textwidth]{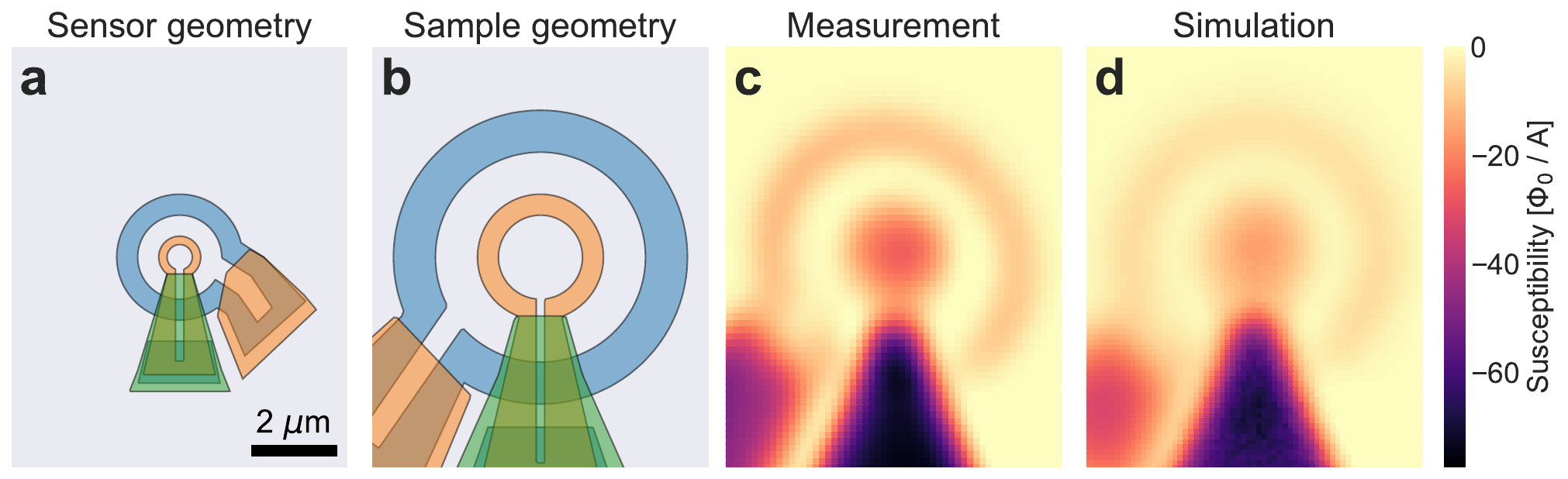}
    \caption{Simulation of a scanning SQUID susceptometry measurement. ({\bf a}) Geometry of the \inline{Device} representing the ``sensor'' SQUID. ({\bf b}) Geometry of the \inline{Device} representing the ``sample'' SQUID. ({\bf c}) Susceptibility of the sample SQUID measured at a temperature of 4.0 K. ({\bf d}) Simulated susceptibility, calculcated using the method described in Section~\ref{section:examples:scanning-squid}. The pickup loop of the sample, which in reality is connected to a SQUID circuit that has a non-linear magnetic response, is modeled as a continuous superconducting loop in the zero-fluxoid (i.e. Meissner) state. The susceptibility signal is defined as $M_\text{PL-FC} - M_\text{PL-FC, no sample}$, where $M_\text{PL-FC, no sample}\approx163\,\Phi_0/\mathrm{A}$ (see Figure~\ref{fig:squid-mutuals}({\bf a}), second column). The scale bar in ({\bf a}) applies to all four panels.}
    \label{fig:squid-susc}
\end{figure*}

Figure~\ref{fig:squid-mutuals}({\bf a}) shows \SuperScreen \inline{Device} models of the field coil and pickup loop region of four types of real SQUID susceptometers, with geometry taken from the layout artwork (GDS) files. In addition to the field coil and pickup loop, there are superconducting shields that limit the amount of magnetic flux that can penetrate the leads connecting the loops to the rest of the circuit (the rest of the circuit is not modeled). There are three relevant layers of superconducting films: the base electrode (BE), which is furthest from the sample contains the field coil; the first wiring layer (W1), which contains the pickup loop and a shield covering the field coil leads; and the second wiring layer (W2), which is closest to the sample and contains a shield covering the pickup loop leads. There are two insulating layers: I1, which separates BE and W1, and I2, which separates W1 and W2.
For the superconducting layers, which are made of Nb, we take the London penetration depth to be $\lambda = 80\,\nm$, and the layer thicknesses correspond to the real device design (see Table 1 of Ref.~\cite{Kirtley2016-gt} and Figure 8 of Ref.~\cite{Kirtley2016-zz}). It is important to note that because $\lambda < d$ for all three layers, the susceptometers are not in the 2D limit in which the model described in Section~\ref{section:model} is technically valid. Nevertheless, as shown in Figure~\ref{fig:squid-mutuals}, the field coil - pickup loop mutual inductances, $M_\mathrm{sim}$, computed by \SuperScreen lie within the range of mutual inductances measured in real devices, $M_\mathrm{exp}$ (taken from Table 1 of Ref.~\cite{Kirtley2016-zz}), for all four sizes of susceptometer, indicating that variation in current density along the thickness of each film is not critical in determining the mutual inductance in this case. Figure~\ref{fig:squid-mutuals}({\bf c}) shows the convergence of $M_\mathrm{PL-FC}$ as a function of solver iteration (see Section~\ref{section:model:multilayer}).

Having established the value of the field coil - pickup loop mutual inductance in the absence of a sample, $M_\text{PL-FC, no sample}$, we can simulate a scanning SQUID susceptometry measurement by calculating $M_\mathrm{PL-FC}$ in the presence of a superconducting sample as a function of the relative position of the pickup loop and the sample, $(x_s, y_s)$. For example, in Figure~\ref{fig:squid-susc} we simulate a SQUID susceptometry measurement of a ``Large'' susceptometer (the sample) measured with a ``Medium'' suceptometer (the sensor). Given a \inline{Device} representing the sensor and a \inline{Device} representing the sample, this calculation is performed in three steps:
\begin{enumerate}
    \item{Simulate the sensor with some known current $I_\mathrm{FC}$ circulating in the field coil to obtain \inline{field_coil_solution} and $M_\text{PL-FC, no sample}$.}
    \item{For a given relative position $(x_s, y_s)$ between the sensor and sample, simulate the sample with an applied field given by \inline{field_coil_solution.field_at_position()} to obtain \inline{sample_solution}.}
    \item{Simulate the sensor again, this time with the applied field given by \inline{sample_solution.field_at_position()}, evaluate the fluxoid $\Phi^f_\mathrm{PL}$ for the hole representing the sensor's pickup loop, and calculate the mutual inductance: $M_\mathrm{PL-FC}(x_s, y_s)=\Phi^f_\mathrm{PL} / I_\mathrm{FC}$.}
\end{enumerate}
These three steps are repeated for every desired $(x_s, y_s)$ to build up a susceptometry image. The susceptibility signal is reported as $M_\text{PL-FC}(x_s, y_s) - M_\text{PL-FC, no sample}$, where values that are more negative indicate a stronger diamagnetic response from the sample. Note that in principle it is possible to combine the sensor and sample into a single \inline{Device} with 6 superconducting layers, however this is impractical because it would require generating a new mesh for each $(x_s, y_s)$ and, as discussed in detail in \ref{appendix:numerics}, the problem scales unfavorably with the number of layers in a device and with the total lateral extent of a device.

\section{Conclusion}
\label{section:conlusion}

The ability to model and visualize screening effects in 2D superconductors and devices constructed from superconducting thin films can help to build intuition about these systems, aid in interpretation of measurement results, and enable optimization of measurement and device design. \SuperScreen is an open-source, user-friendly, portable, and efficient tool that solves this problem. Applications of the package include calculating self- and mutual-inductance in planar and multi-planar superconducting circuits, and modeling the magnetic interaction between superconducting samples and superconducting sensors such as scanning SQUID susceptometers~\cite{Kirtley2016-zz}.

There are several important limitations to the applicability of \SuperScreen and the matrix inversion method on which it is based~\cite{Brandt2004-ew,Brandt2005-wj}. First, strictly speaking all superconducting films should be in the 2D limit, with London penetration depth $\lambda$ large compared to the film thickness $d$, such that the current density is approximately constant along the thickness of the film. There are cases where the model reproduces experimental results despite violation of this condition (e.g. the calculations and in Refs.~\cite{Kirtley2016-zz,Kirtley2016-gt} and Section~\ref{section:examples:scanning-squid}), but care must be taking in interpreting results in these cases. Second, the model assumes that all superconducting films behave linearly and without dissipation, and that the applied magnetic field and current density are well below the critical field and critical current density of all films in a device. Third, \SuperScreen does not support ``terminal currents,'' i.e. currents flowing in one terminal of a device and out another terminal. This means that inductance calculations are limited to structures with holes, in which all applied currents are circulating currents associated with trapped flux. Terminal currents can, however, be included in stream function-based models by setting appropriate boundary conditions~\cite{Khapaev1997-kw,Khapaev2001-xq,Khapaev2001-pw, Khapaev2002-ev, Muller2021-ci}. An extension to the model described above that treats the magnetic response of a superconducting ring interrupted by two Josephson junctions (i.e. a SQUID) with trapped vortices and terminal currents is given in Ref.~\cite{Clem2005-ye}. Finally, care should be taken to ensure that for a given model the mesh is of sufficient density that, to within the desired precision, the results of simulations do not depend on mesh size (see \ref{appendix:numerics}).

Potential improvements to \SuperScreen include: support for terminal currents as discussed above, automated determination of solution convergence for models with multiple layers, more sophisticated mesh generation (e.g. automated local mesh refinement based on device geometry or adaptive mesh refinement based on solution convergence), integration with standard integrated circuit layout software or file formats, and further numerical optimization, including GPU acceleration.

\section{Acknowledgements}
\label{section:acknowledgements}
We acknowledge useful discussions with John R. Kirtley. This work is supported by the Department of Energy, Office of Science, Basic Energy Sciences, Materials Sciences and Engineering Division, under Contract DE-AC02-76SF00515. Some of the computing for this project was performed on the Sherlock cluster. We would like to thank Stanford University and the Stanford Research Computing Center for providing computational resources and support that contributed to these research results.



\appendix

\section{Existing tools}
\label{appendix:other-tools}

Here, we briefly describe existing software tools for modeling the magnetic response of superconducting devices, most of which are specifically designed for inductance extraction for superconducting integrated circuits. FastHenry, a widely-used 3D (normal metal) inductance extraction tool from MIT~\cite{Kamon1994-ck}, has been extended to support superconducting elements~\cite{wrcad, XicTools}. This modified version of FastHenry has been used for inductance extraction in the commercial software InductEx~\cite{Fourie2011-wl, Fourie2012-gv, Jackman2016-mf}. While FastHenry is open-source, it is written in C and must be compiled for a specific computer architecture and operating system. FastHenry executables compiled for several common operating systems are available as part of the open-source XicTools superconducting integrated circuit design suite from Whiteley Research, Inc~\cite{wrcad, XicTools}. A 2D London-Maxwell approach based on a scalar stream function, much like the approach used by \SuperScreen~\cite{Brandt2004-ew, Brandt2005-wj}, forms the basis of the 3D-MLSI software package~\cite{Khapaev1997-kw, Khapaev2001-xq, Khapaev2001-pw, Khapaev2002-ev}, which is also written in C and is not open-source. For a thorough historical overview and comparison of inductance extraction tools, see Ref.~\cite{Gaj1999-ls}. For a more recent overview of superconducting inductance extraction tools see Ref.~\cite{Tolpygo2021-jz}, particularly references [8--25] therein.

\SuperScreen is not intended primarily as an inductance extraction tool and cannot be used to extract the inductance of superconducting transmission lines (a common superconducting circuit design task~\cite{Gaj1999-ls,Khapaev1996-uu,Tolpygo2021-jz}) due to the lack of support for ``terminal currents.'' Nevertheless, it is instructive to compare \SuperScreen to a state of the art commercial inductance extraction tool, InductEx~\cite{Fourie2011-wl, Fourie2012-gv, Jackman2016-mf}, on a problem that can be solved using both tools. One such problem, provided as an example in the InductEx user manual, is the self-inductance $L$ of a square washer with outer dimension $30\,\um$, inner dimension $10\,\um$, London penetration depth $\lambda=0.24\,\um$, and film thickness $d=0.20\,\um$, for which the InductEx solution reports a ``design'' value of $L=20\,\mathrm{pH}$ and an ``extracted'' value of $L=20.0956\,\mathrm{pH}$~\cite{inductex_user_manual_2020}. Code Block~\ref{code:washer} demonstrates how to solve this problem using \SuperScreen. The inductance extracted using \SuperScreen, $L=19.91\,\mathrm{pH}$, is within 1\% of the inductance extracted using InductEx.

\begin{code-onecol}
\begin{minted}[fontsize=\small]{python}
import superscreen as sc
from superscreen.geometry import box

layer = sc.Layer(
    "base",
    london_lambda=0.24,
    thickness=0.2,
)
film = sc.Polygon(
    "washer",
    layer="base",
    points=box(30, points_per_side=50),
)
hole = sc.Polygon(
    "hole",
    layer="base",
    points=box(10, points_per_side=250),
)
device = sc.Device(
    "washer",
    layers=[layer],
    films=[film],
    holes=[hole],
    length_units="um",
)
device.make_mesh(min_points=3000)
# See Equation 19.
M = device.mutual_inductance_matrix()
L = M[0, 0].to("pH")
print(f"Inductance = {L:~.4fP}")
# Output: Inductance = 19.9100 pH
\end{minted}
\captionof{listing}{\SuperScreen script to calculate the self-inductance of a square washer with outer dimension $30\,\um$, inner dimension $10\,\um$, London penetration depth $\lambda=0.24\,\um$, and film thickness $d=0.20\,\um$. The extracted inductance $L=19.91\,\mathrm{pH}$ differs from the InductEx ``design'' and ``extracted'' values by -0.45\% and -0.92\% respectively~\cite{inductex_user_manual_2020}.}
\label{code:washer}
\end{code-onecol}

\section{Mesh Laplace and gradient operators}
\label{appendix:laplace}

The definitions of the Laplace operator $\mathbf{\nabla}^2$ (also called the Laplace-Beltrami operator) and the gradient operator $\vec{\nabla}=(\nabla_x, \nabla_y)^T$ deserve special attention, as these two operators reduce the problem of solving a partial differential equation to the problem of solving a matrix equation~\cite{Reuter2009-hr}.  Given a mesh consisting of $p$ vertices and $t$ triangles, and a scalar field $f(x, y)$ represented by a $p\times 1$ vector $\mathbf{f}$ containing the values of the field at the mesh vertices, the goal is to construct matrices $\nabla^2$ and $\vec{\nabla}=(\nabla_x, \nabla_y)^T$ such that the matrix products $\nabla^2\mathbf{f}$ and $\vec{\nabla}\mathbf{f}$ approximate the Laplacian $\left(\frac{\partial^2f}{\partial x^2}+\frac{\partial^2f}{\partial y^2}\right)$ and the gradient $\left(\frac{\partial f}{\partial x}, \frac{\partial f}{\partial y}\right)$  of $f(x, y)$ at the mesh vertices.

As described in Ref.~\cite{Crane_Vouga_2014}, the Laplace operator $\mathbf{\nabla}^2$ for a mesh is defined in terms of two matrices, the mass matrix $\mathbf{M}$ and the
weak Laplacian matrix $\mathbf{L}$: $\mathbf{\nabla}^2 = \mathbf{M}^{-1}\mathbf{L}$. In a 2D mesh, the mass matrix $\mathbf{M}$ gives an effective area to each vertex in the mesh. Here we use a ``lumped" mass matrix, which is diagonal with elements $M_{ii} = \frac{1}{3}\sum_{t\in\mathcal{N}(i)}\mathrm{area}(t)$,
where $\mathcal{N}(i)$ is the set of triangles $t$ adjacent to vertex $i$. The weak Laplacian matrix $\mathbf{L}$ is defined in terms of a symmetric weight matrix $\mathbf{W}$, which assigns a weight to every edge in the mesh. $\mathbf{W}$ may be defined in a number of ways:

\begin{enumerate}
    \item{
        Uniform weighting: In this case, $\mathbf{W}$ is simply the adjacency matrix for the mesh vertices:
        $$
            W_{ij} =
            \begin{cases}
                0&\text{if }i=j\\
                1&\text{if }i\text{ is adjacent to }j\\
                0&\text{otherwise}
            \end{cases}
        $$
    }
    \item{
        Inverse-Euclidean weighting: Each edge is weighted by the inverse of its length: $|\vec{r}_i-\vec{r}_j|^{-1}$, where $\vec{r}_i$ is the position of vertex $i$.
        $$
            W_{ij} =
            \begin{cases}
                0&\text{if }i=j\\
                |\vec{r}_i-\vec{r}_j|^{-1}&\text{if }i\text{ is adjacent to }j\\
                0&\text{otherwise}
            \end{cases}
        $$
    }
    \item{
        Half-cotangent weighting: Each edge is weighted by the half the sum of the cotangents of the two angles opposite to it.
        $$
            W_{ij} =
            \begin{cases}
                0&\text{if }i=j\\
                \frac{1}{2}\left(\cot\alpha_{ij}+\cot\beta_{ij}\right)&\text{if }i\text{ is adjacent to }j\\
                0&\text{otherwise}
            \end{cases}
        $$
    }
\end{enumerate}

By default, \SuperScreen uses half-cotangent weighting. The Laplacian matrix $\mathbf{L}$ is defined in terms of the weight matrix $\mathbf{W}$: $L_{ij} = W_{ij} - \delta_{ij}\sum_{\ell}W_{i\ell}$. Finally, the Laplace operator is given by $\mathbf{\nabla}^2 = \mathbf{M}^{-1}\mathbf{L}$.

We construct the two $p\times p$ gradient matrices $\nabla_x$ and $\nabla_y$, using the ``average gradient on a star" (or AGS) approach~\cite{Mancinelli2019-ci}. Briefly, we first construct two $t\times p$ matrices, $\vec{\nabla}_t=(\nabla_{t,x}, \nabla_{t,y})^T$ using ``per-cell linear estimation'' (or PCE)~\cite{Mancinelli2019-ci}, where $\nabla_{t,x}\mathbf{f}$ maps the field values at the vertices $\mathbf{f}$ to an estimate of the $x$-component of the gradient at the triangle centroids (centers-of-mass). The matrices $\nabla_x$ and $\nabla_y$ are then computed by, for each vertex $i$, taking the weighted average of $\nabla_{t,x}$ and $\nabla_{t,y}$ over adjacent triangles $t\in\mathcal{N}(i)$, with weights given by the angle between the two sides of the triangle adjacent to the vertex. The resulting $\vec{\nabla}=(\nabla_x, \nabla_y)^T$ is a $2\times p\times p$ stack of matrices defined such that $\vec{\nabla}\mathbf{f}$ produces a $2\times p$ matrix representing the gradient of $f(x, y)$ at the mesh vertices, with the first and second rows of $\vec{\nabla}\mathbf{f}$ containing the $x$ and $y$ components of the gradient, respectively.

\section{Spatially inhomogeneous $\lambda$}
\label{appendix:inhomogeneous}

The London equation (Equation~\ref{eq:london}) is valid only under the assumption that the London penetration depth $\lambda$, a proxy for the superfluid density, is constant as a function of position~\cite{Tinkham2004-zn}. In cases where the superfluid density varies as a function of position, Ginzburg-Landau theory provides a more accurate description of the magnetic response of the system. Nevertheless, in an effort to model inhomogeneous superconductors using London theory, one can write out the ``inhomogeneous second London equation'' for a superconductor with spatially-varying London penetration depth $\lambda(\vec{r})$~\cite{Cave1986-js, Kogan2011-zn}:

\begin{align}
\begin{split}
    \label{eq:london_inhomogeneous_3d}
    \vec{H}(\vec{r})&=-\vec{\nabla}\times\left(\lambda^2(\vec{r})\vec{j}(\vec{r})\right)\\
    &=-\left(\lambda^2(\vec{r})\vec{\nabla}\times\vec{j}(\vec{r}) + \vec{\nabla}\lambda^2(\vec{r})\times\vec{j}(\vec{r})\right).
\end{split}
\end{align}

In the 2D limit, i.e. a thin film with thickness $d\ll\lambda(x, y)$ lying parallel to the $x-y$ plane carrying sheet current density $\vec{J}(x, y)=\vec{j}(\vec{r})\cdot d$, we have:
\begin{align}
\begin{split}
    \label{eq:london_inhomogeneous_2d}
    \vec{H}(x, y)&=-\vec{\nabla}\times(\Lambda\vec{J})\\
    &=-\left(\Lambda\vec{\nabla}\times\vec{J}+\vec{\nabla}\Lambda\times\vec{J}\,\right)\\
    &=\left(\Lambda\nabla^2g+\vec{\nabla}\Lambda\cdot\vec{\nabla}g\right)\hat{z},
\end{split}
\end{align}
where $\Lambda=\Lambda(x, y)$, $g=g(x, y)$, $\vec{\nabla}=\left(\frac{\partial}{\partial x},\frac{\partial}{\partial y}\right)$, and $\vec{J}=\vec{J}(x, y)=\vec{\nabla}\times(g\hat{z})$.

If one defines an inhomogeneous effective penetration depth $\Lambda(x, y)$ in a \SuperScreen model, Equation~\ref{eq:london_inhomogeneous_2d}, rather than Equation~\ref{eq:london}, is solved numerically as follows. For a mesh with $p$ vertices, the effective penetration depth is represented by a $p\times 1$ vector $\mathbf{\Lambda}$. Equations~\ref{eq:Heff_sub_num} and \ref{eq:full_stream} are updated according to:
\begin{align}
\begin{split}
    \label{eq:Heff_inhomogeneous}
    &\mathbf{Q}.\mathbf{w}^T-\Lambda\mathbf{\nabla}^2\to\\
    &\mathbf{Q}.\mathbf{w}^T-\mathbf{\Lambda}^T.\mathbf{\nabla}^2-\vec{\nabla}\mathbf{\Lambda}\cdot\vec{\nabla}
\end{split}
\end{align}

The notation $\vec{\nabla}\mathbf{f}\cdot\vec{\nabla}$ indicates an inner (dot) product over the two spatial dimensions, resulting in a $p\times p$ matrix such that $(\vec{\nabla}\mathbf{f}\cdot\vec{\nabla})\mathbf{g}$ computes $(\vec{\nabla}f(x, y))\cdot(\vec{\nabla}g(x, y))$ (see~\ref{appendix:laplace}).

Note that, unlike Equation~\ref{eq:london} in which $\Lambda$ is assumed to be constant as a function of position $\pvec{r}$, solutions to Equation~\ref{eq:london_inhomogeneous_2d} do not necessarily satisfy the fluxoid quantization condition $\Phi^f_S=0$ for simply-connected superconducting regions $S$ in which $\vec{\nabla}\Lambda(\pvec{r})\neq0$ , where
\begin{equation}
	\Phi^f_S=\underbrace{\int_S\mu_0H_z(\pvec{r})\,\mathrm{d}^2r}_{\text{``flux part''}} + \underbrace{\oint_S\mu_0\Lambda(\pvec{r})\vec{J}(\pvec{r})\cdot\mathrm{d}\pvec{r}}_{\text{``supercurrent part''}}.
\end{equation}
 
\section{Numerical considerations}
\label{appendix:numerics}
\begin{figure}[t]
    \centering
    \includegraphics[width=\linewidth]{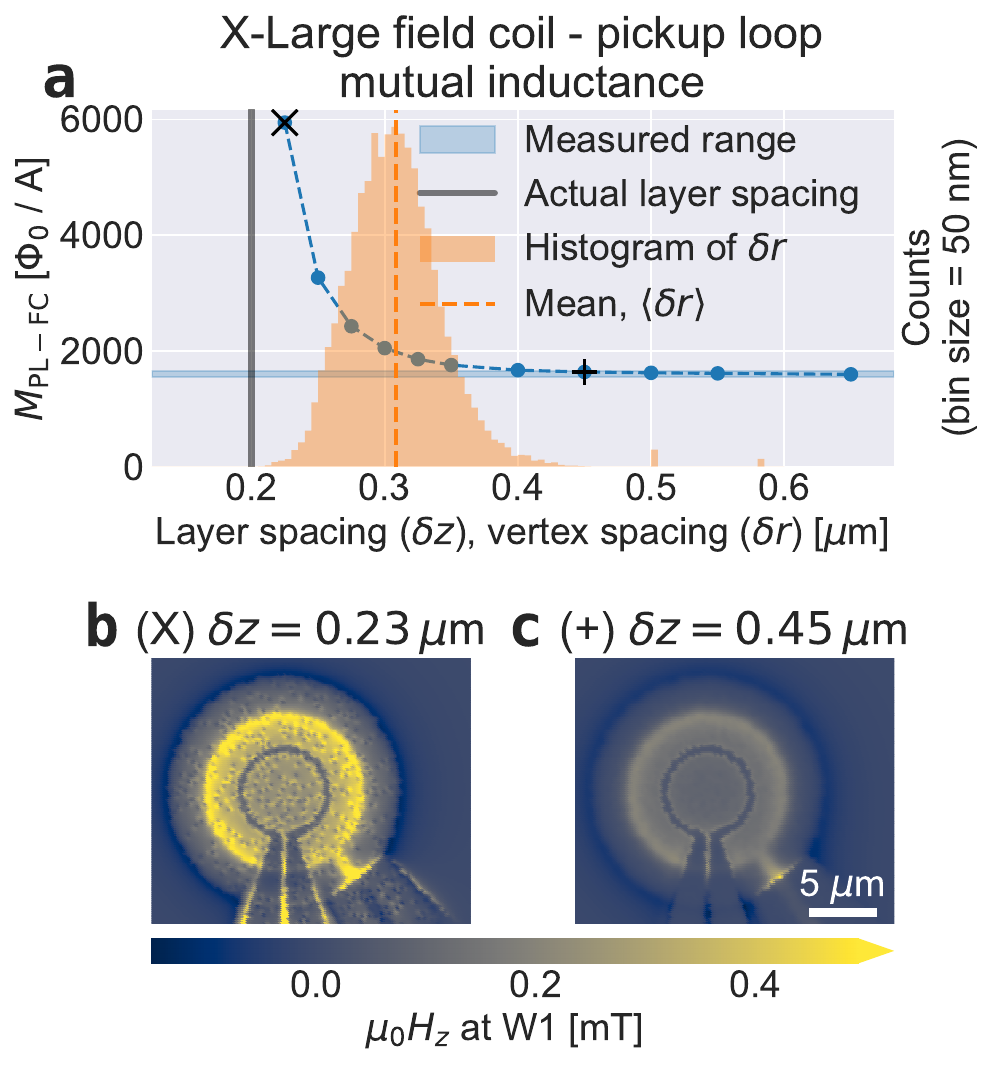}
    \caption{\SuperScreen results may be unreliable for multi-layer models where the spacing between layers is smaller than the typical vertex-to-vertex distance, $\delta r$. ({\bf a}) Field coil - pickup loop mutual inductance for the largest SQUID suscepometer modeled in Section~\ref{section:examples:scanning-squid} as a function of the minimum vertical spacing between layers, $\delta z$. The simulated mutual inductance is shown with blue circles, and a histogram of the mesh vertex-to-vertex distances $\delta r$ is shown in orange (arb. $y$ axis units). When $\delta z\lesssim\delta r$, the model significantly overestimates the mutual inductance. ({\bf b}) and ({\bf c}): Out-of-plane magnetic field $\mu_0H_z$ evaluated at the plane of the pickup loop (W1 layer) for ({\bf b}) the nominal layer spacing with $\delta z<\delta r$ (indicated with a black $\times$ in ({\bf a})) and ({\bf c}) $\delta z > \delta r$ (indicated with a black $+$ in ({\bf a})). For $\delta r\lesssim\delta z$, the magnetic field calculated using Eq.~\ref{eq:hz-from-kernel} resembles that of a discrete set of isolated dipoles rather than a continuous sheet of current. Note that ({\bf b}) and ({\bf c}) share the same color scale, which is saturated in ({\bf b}).}
    \label{fig:xlarge-dz-dr}
\end{figure}

While the numerical method described in Section~\ref{section:implementation} is generally quite robust, it can break down for certain extreme geometries. The stream function $g(x, y)$ represents the local magnetization or density of infinitesimal current loops. The $z$-component of the magnetic field at position $\vec{r}=(x, y, z)$ from a film $F$ with stream function $g$ lying in a plane parallel to the $x-y$ plane at vertical position $z'$ is given by (see Equations~\ref{eq:field_from_kernel} and \ref{eq:kernels}):
\begin{align}
\label{eq:hz-from-kernel}
\begin{split}
    H_z(\vec{r}) &= \int_F Q_z(\vec{r},\pvec{r}')g(x', y')\,\mathrm{d}^2r',\,\text{where}\\
    Q_z(\vec{r}, \pvec{r}') &=  \frac{2(z-z')^2-\rho^2}
            {4\pi[(z-z')^2+\rho^2]^{5/2}}
\end{split}
\end{align}
and $\rho=\sqrt{(x-x')^2+(y-y')^2}$. Eq.~\ref{eq:hz-from-kernel} is exact for a continuous stream function $g$. However, the discretized version of Eq.~\ref{eq:hz-from-kernel}, in which the double integral over the film area $F$ is replaced by a sum over triangular mesh elements, is only valid if $\delta z = z-z'$, the vertical distance between the film and the point at which the field is being evaluated, is large compared to the typical distance $\delta r$ between vertices in the mesh representing the film. For $z-z'=\delta z\lesssim\delta r$, the field $H_z(\vec{r})$ resembles that of a discrete set of isolated dipoles located at the mesh vertex positions, rather than that of a continuous sheet of current (see Figure~\ref{fig:xlarge-dz-dr}({\bf b})). This can lead to unphysical results when evaluating the field very close to the surface of a film (for example using \inline{Solution.field_at_position()}), or when solving models involving multi-layer structures where the vertical spacing between layers is much smaller than the lateral extent of the films, in which case the iterative calculation (Section~\ref{section:model:multilayer}) may not converge.

The limitation described above can be seen in the model of the largest SQUID susceptometer described in Section~\ref{section:examples:scanning-squid}, which has a field coil inner radius of $6\,\um$ and a total modeled area of roughly $600\,\um^2$. As shown in Figure~\ref{fig:xlarge-dz-dr}, \SuperScreen significantly overestimates the mutual inductance between the field coil and pickup loop when $\delta z <\delta r$ because, in that case, the discretized version of Eq.~\ref{eq:hz-from-kernel} does not correctly compute the magnetic field due the stream functions of the superconducting layers. For the sake of computing mutual inductance as in Section~\ref{section:examples:scanning-squid}, it is physically reasonable to artificially increase the vertical spacing between layers such that $\delta z\geq \delta r$ because we expect the magnetic field at the pickup loop, a vertical distance $\delta z$ away from the center of the field coil, to fall off roughly as $\left(\delta z^2 + R^2_\mathrm{FC}\right)^{-3/2}$, where $R_\mathrm{FC}\approx 6\,\um\gg\delta z$. However, in situations where $\delta z$ is a critical dimension (in the sense that increasing it would invalidate the physical model), one's only option is to decrease $\delta r$ by increasing the density of the mesh.

\begin{figure}[t]
    \centering
    \includegraphics[width=\linewidth]{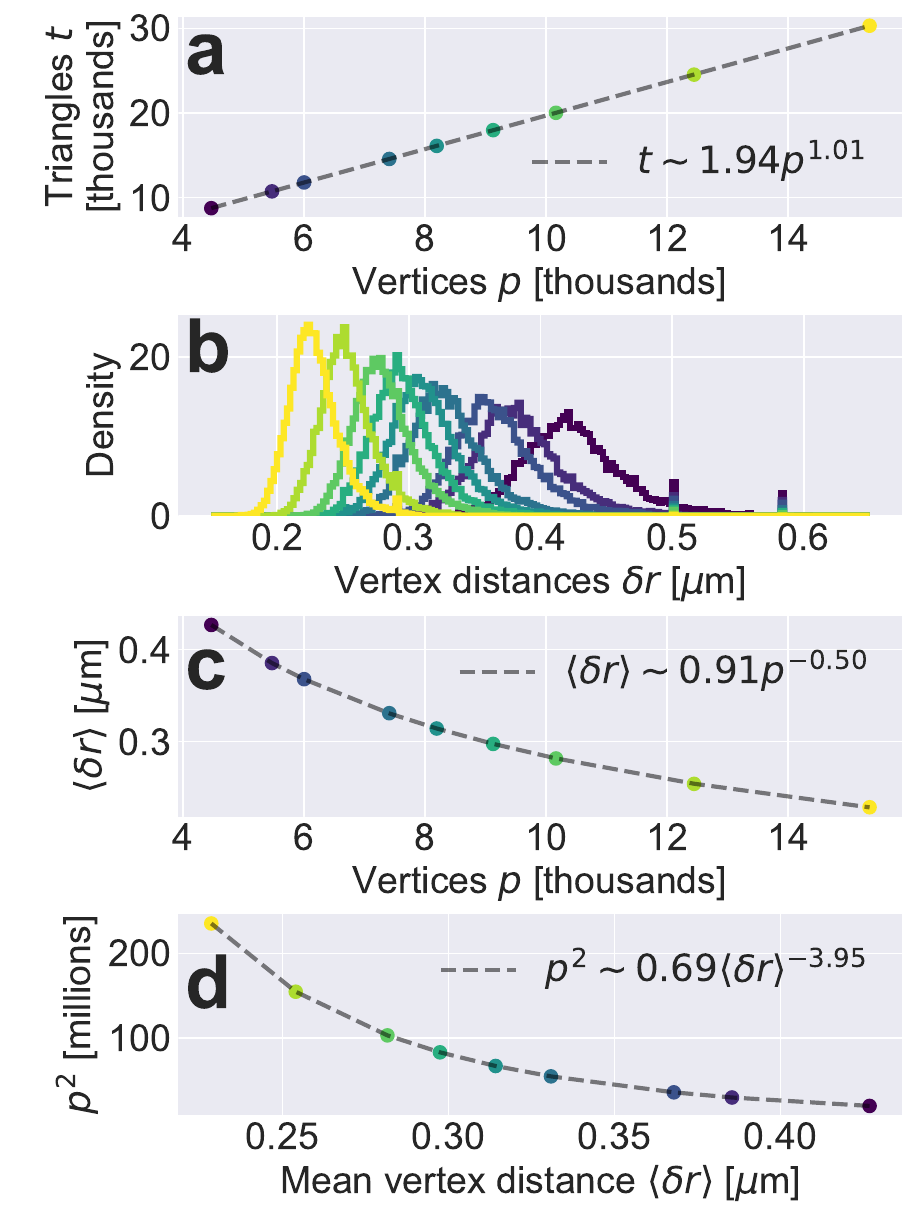}
    \caption{Scaling of kernel matrix size $p^2$ with mesh vertex spacing $\delta r$ for a mesh with $p$ vertices and $t$ triangles for the device shown in Figure~\ref{fig:xlarge-dz-dr}. ({\bf a}) Typically the number of triangle $t$ in the mesh is close to twice the number of vertices $p$. ({\bf b}) Normalized histograms of vertex distances $\delta r$. The histogram bins are given by \inline{numpy.linspace(0.15, 0.65, 201)}. ({\bf c}) The mean vertex-to-vertex distance $\langle\delta r\rangle$ scales roughly as $p^{-1/2}$. ({\bf d}). The size of each $p\times p$ kernel matrix (of which where are $\binom{L}{2}$ for a device with $L$ layers) scales roughly as $\langle\delta r\rangle^{-4}$. Note that in each row the colors correspond to the number of vertices $p$ in the mesh, as indicated in ({\bf a}), with mesh size increasing from dark to light colors.}
    \label{fig:scaling}
\end{figure}

For this reason, mutli-layer structures with closely-spaced layers are the most challenging class of problem to solve. The iterative method used described in Section~\ref{section:model:multilayer} is memory-intensive for models with a large mesh (many vertices $p$ and triangles $t$, with typically $t\approx 2p$) and/or many layers $L$, because the average distance between vertices decreases slowly with increasing number of vertices, $\langle\delta r\rangle\sim p^{-1/2}$, whereas the size of the (dense floating-point) $p\times p$ matrix that represents the dipole kernel $Q_z(\vec{r},\pvec{r}')$ in Eq.~\ref{eq:hz-from-kernel} increases as $p^2$. The end result is that the memory footprint of kernel matrix scales roughly as $\langle\delta r\rangle^{-4}$, so decreasing the mean distance between vertices by a factor of 2 increases the memory required by a factor of roughly 16 (see Figure~\ref{fig:scaling}).

Furthermore, for a model with $L$ layers there are $\binom{L}{2} = L(L-1)/2$ such kernel matrices needed for each iteration of the calculation outlined in Section~\ref{section:model:multilayer}. In \inline{superscreen.solve()}, these $\binom{L}{2}$ matrices are computed during the first iteration and then cached in memory for use in subsequent iterations. One can force the kernel matrices to be cached to disk if they would otherwise occupy too large a fraction of the available system memory using the \inline{cache_kernel_memory_cutoff} argument, but this comes at a significant performance cost. In many cases, one can use lower-precision floating point numbers (e.g. using 32-bit single-precision floats instead of the default 64-bit double-precision floats by setting \inline{device.solve_dtype = "float32"}) to reduce memory requirements without significantly impacting solution accuracy.

\section{Parallel processing}
\label{section:parallel}
As discussed above, one can solve many models involving the same \inline{Device} in parallel across multiple CPUs using the \inline{superscreen.solve_many()} function. There are two methods available for process-based parallelism in \SuperScreen: \inline{parallel_method="multiprocessing"}, which uses the \inline{multiprocessing} package from the Python standard library, and \inline{parallel_method="ray"}, which uses the third-party distributed computing framework \inline{Ray}\footnote{Note that at the time of writing, \inline{Ray} support for Windows is experimental and under active development.}~\cite{Moritz2018-mt,ray-docs}. Both approaches utilize shared memory so that only a single copy is made of the large arrays required to solve a \inline{Device} (the mesh, kernel matrix $\mathbf{Q}$, Laplace operator $\mathbf{\nabla}^2$, etc.), rather than \inline{num_cpus} copies, where \inline{num_cpus} is the number of worker processes.

\begin{code-onecol}
\begin{minted}[fontsize=\small]{python}
parallel_method = "multiprocessing"
# parallel_method = "ray"

# Specify number of worker processes:
num_cpus = 4
# Or automatically use all
# available physical CPUs:
# num_cpus = None

_ = superscreen.solve_many(
    device=device,
    parallel_method=parallel_method,
    num_cpus=num_cpus,
    **solve_kwargs,
)
\end{minted}
\captionof{listing}{Utilizing process-based parallelism in \SuperScreen given a \inline{superscreen.Device} and solver options stored in a dictionary \inline{solve_kwargs}.}
\label{code:parallel}
\end{code-onecol}

There are three ways to invoke \inline{Ray} from \SuperScreen when running on a single machine, e.g. a multi-core CPU. The first is to simply pass the keyword argument \inline{parallel_method="ray"} when calling \inline{solve_many()} (see Code Block~\ref{code:parallel}). This will automatically create a \inline{Ray} cluster using (by default) all available physical CPU cores, solve the models in parallel, and then shut down the cluster before returning. The second method is to manually create a \inline{Ray} cluster using the \inline{Ray} Python application programming interface (API) prior to calling \inline{solve_many(..., parallel_method="ray")}, as demonstrated in Code Block~\ref{code:ray-python}. The third method is to start a \inline{Ray} cluster using the command line interface (CLI), then connect to the existing cluster using the Python API prior to calling \inline{solve_many(..., parallel_method="ray")}, as demonstrated in Code Block~\ref{code:ray-cli}. One of the latter two methods should be used for finer control over the \inline{Ray} cluster. For example, if calling \inline{solve_many()} many times in a single session, one can use these methods to avoid the overhead of starting and stopping a \inline{Ray} cluster multiple times.

Running \inline{superscreen.solve_many()} in parallel across multiple nodes in a computing cluster is a simple extension to the method outlined in Code Block~\ref{code:ray-cli}, although the specifics depend upon the infrastructure of the cluster, e.g. job management software. See the ``Multi-Node Ray" section of the \inline{Ray} documentation for more details~\cite{ray-docs}.

\begin{code}
\begin{minted}[fontsize=\small]{python}
# Assume that we have already created a Device and put all other inputs
# to superscreen.solve_many() into a dictionary called other_kwargs.
import psutil
import ray

# Specify the number of CPUs/cores to allocate.
num_cpus = 3
# Use at most N processes for a machine with N physical CPUs.
num_cpus = min(num_cpus, psutil.cpu_count(logical=False))

# Start a ray cluster
ray.init(num_cpus=num_cpus)
# Solve the models
solutions, paths = superscreen.solve_many(
    parallel_method="ray",
    **other_kwargs,
)
# Potentially call solve_many() again using the same ray cluster...
# Finally, shutdown the ray cluster
ray.shutdown()
\end{minted}
\captionof{listing}{Starting and stopping \inline{Ray} outside of \inline{superscreen.solve_many()} using the Python API. See the ``API and Package Reference" section of the \inline{Ray} documentation for additional options in \inline{ray.init()}~\cite{ray-docs}.}
\label{code:ray-python}
\end{code}

\begin{code}
\begin{minipage}{1.0\textwidth}
\begin{minted}[fontsize=\small]{bash}
# Start a ray cluster from the command line, e.g. bash
ray start --head --num-cpus=3
\end{minted}

\begin{minted}[fontsize=\small]{python}
# Assume that we have already created a Device and put all other inputs
# to superscreen.solve_many() into a dictionary called other_kwargs.
import ray
# Connect to the existing ray cluster.
# If more than one ray cluster is running, specify
# which to connect to using address="{ip}:{port}".
ray.init(address="auto")
# Solve the models.
solutions, paths = superscreen.solve_many(
    parallel_method="ray",
    **other_kwargs,
)
# Potentially call solve_many() again using the same ray cluster...
\end{minted}

\begin{minted}[fontsize=\small]{bash}
# Shut down the ray cluster from the command line
ray stop
\end{minted}
\captionof{listing}{Starting and stopping \inline{Ray} outside of \inline{superscreen.solve_many()} using the command line interface. See the \inline{Ray} documentation for additional options in \inline{ray start}~\cite{ray-docs}.}
\label{code:ray-cli}
\end{minipage}
\end{code}



\bibliographystyle{elsarticle-num}
\bibliography{references}







\end{document}